\documentclass[longauth]{aa} %blank page appears between abstract and intro unless longauth is used  

\usepackage{graphicx}
\usepackage{txfonts}
\usepackage{xcolor}
\usepackage{multirow}
\usepackage{array}
\usepackage{siunitx}
\usepackage{booktabs}
\usepackage{amsmath}
\usepackage{multirow} %added
\usepackage{hyperref}
\begin{document} 
\newcommand{\sgr}{SGR\,0501+4516}

   \title{The infrared counterpart and proper motion of magnetar SGR\,0501$+$4516}

   \subtitle{}

   \author{A. A. Chrimes
          \inst{1}\fnmsep\inst{2}\thanks{ESA Research Fellows}\fnmsep\thanks{Authors contributed equally},\ 
          A. J. Levan\inst{2}\fnmsep\inst{3}$^{\star \star}$,\ 
          J. D. Lyman\inst{3},\ 
          A. Borghese\inst{4}$^{\star}$,\
          V. S. Dhillon\inst{5,6},\
          P. Esposito\inst{7},\
          M. Fraser\inst{8},\
          A. S. Fruchter\inst{9},\
          D. G\"{o}tz\inst{10},\
          R. A. Hounsell\inst{11,12},\
          G. L. Israel\inst{13},\
          C. Kouveliotou\inst{14,15},\
          S. Mereghetti\inst{16},\
          R. P. Mignani\inst{16,17},\
          R. Perna\inst{18},\ 
          N. Rea\inst{19,20},\ 
          I. Skillen\inst{21},\ 
          D. Steeghs\inst{3},\
          N. R. Tanvir\inst{22},\
          K. Wiersema\inst{23},\ 
          N. J. Wright\inst{24},\
          S. Zane\inst{25}\
          }

   \institute{European Space Agency (ESA), European Space Research and Technology Centre (ESTEC), Keplerlaan 1, 2201 AZ Noordwijk, the Netherlands \\
              \email{ashley.chrimes@esa.int}
         \and
             Department of Astrophysics/IMAPP, Radboud University, PO Box 9010, 6500 GL Nijmegen, The Netherlands 
         \and
            Department of Physics, University of Warwick, Gibbet Hill Road, CV4 7AL Coventry, United Kingdom 
         \and
            European Space Agency (ESA), European Space Astronomy Centre (ESAC), Camino Bajo del Castillo s/n, 28692 Villanueva de la Cañada, Madrid, Spain
         \and
            Astrophysics Research Cluster, School of Mathematical and Physical Sciences, University of Sheffield, Sheffield, S3 7RH, United Kingdom
         \and
            Instituto de Astrof\'{\i}sica de Canarias, E-38205 La Laguna, Tenerife, Spain
         \and
            Scuola Universitaria Superiore IUSS Pavia, Palazzo del Broletto, piazza della Vittoria 15, I-27100 Pavia, Italy
         \and
            School of Physics, University College Dublin, L.M.I. Main Building, Beech Hill Road, Dublin 4, D04 P7W1, Ireland
         \and
            Space Telescope Science Institute, 3700 San Martin Drive, Baltimore, MD 21218, USA
         \and
            AIM-CEA/DRF/Irfu/D{\'e}partement d’Astrophysique, CNRS, Universit{\'e} Paris-Saclay, Universit{\'e} de Paris Cit{\'e}, Orme des Merisiers, F-91191 Gif-sur-Yvette, France
         \and
            University of Maryland, Baltimore County, Baltimore, MD 21250, USA
         \and
            NASA Goddard Space Flight Center, Greenbelt, MD 20771, USA
         \and
            INAF–Osservatorio Astronomico di Roma, via Frascati 33, I-00078 Monte Porzio Catone, Italy
         \and
            Department of Physics, The George Washington University, Corcoran Hall, 725 21st St NW, Washington, DC 20052, USA
         \and
            GWU/Astronomy, Physics and Statistics Institute of Sciences (APSIS)
         \and
            INAF - Istituto di Astrofisica Spaziale e Fisica Cosmica Milano, via E. Bassini 15, 20133, Milano, Italy
         \and
            Janusz Gil Institute of Astronomy, University of Zielona G\'ora, ul Szafrana 2, 65-265, Zielona G\'ora, Poland
        \and
            Department of Physics and Astronomy, Stony Brook University, Stony Brook, NY 11794-3800, USA
        \and
            Institute of Space Sciences (ICE, CSIC), Campus UAB, Carrer de Can Magrans s/n, E-08193 Barcelona, Spain
        \and
            Institut d’Estudis Espacials de Catalunya (IEEC), 08860, Castelldefels (Barcelona), Spain
        \and
            Isaac Newton Group of Telescopes, Apartado de Correos 321, E-38700 Santa Cruz de La Palma, Canary Islands, Spain
        \and
            School of Physics and Astronomy, University of Leicester, University Road, Leicester, LE1 7RH, United Kingdom
        \and
            Centre for Astrophysics Research, University of Hertfordshire, College Lane, Hatfield, AL10 9AB, United Kingdom
        \and 
            Astrophysics Research Centre, Keele University, Keele ST5 5BG, United Kingdom
        \and 
            Mullard Space Science Laboratory, University College London, Holmbury St Mary, Dorking, Surrey RH5 6NT, UK
             }

   \date{Received September 15, 1996; accepted March 16, 1997}

% \abstract{}{}{}{}{} 
% 5 {} token are mandatory
 
  \abstract
   {}
   {Soft gamma repeaters (SGRs) are highly magnetised neutron stars (magnetars) notable for their gamma-ray and X-ray outbursts. In this paper, we use near-infrared (NIR) imaging of \sgr\ in the days, weeks, and years after its 2008 outburst to characterise the multi-wavelength emission, and to obtain a proper motion from our long temporal baseline observations.}
   {We present short and long term monitoring of the infrared counterpart of SGR\,0501+4516, and a measurement of its proper motion. Unlike most magnetars, the source has only moderate foreground extinction with minimal crowding. Our observations began only $\sim 2$ hours after the first activation of \sgr\ in August 2008, and continued for $\sim4$ years, including two epochs of Hubble Space Telescope ({\em HST}) imaging. The proper motion constraint is improved by a third {\em HST} epoch 10 years later.}
   %{We present both short and long term monitoring observations of the infrared counterpart of magnetar SGR\,0501+4516, and a measurement of its proper motion. Unlike most known magnetars, the source has only moderate foreground extinction with minimal crowding. Our observations began only $\sim 2$ hours from the first activation of \sgr\ in August 2008, and monitoring continued for $\sim4$ years, including two epochs of Hubble Space Telescope ({\em HST}) imaging. The proper motion constraint is improved by the addition of a third {\em HST} epoch some 10 years later.}
   {The near-infrared and X-rays faded slowly during the first week, thereafter following a steeper power-law decay. The behaviour is satisfactorily fit by a broken power-law. Three epochs of {\em HST} imaging with a 10-year baseline allow us to determine a quiescent level, and to measure a proper motion of $\mu = 5.4 \pm 0.6 $ mas\,yr$^{-1}$. This corresponds to a low transverse peculiar velocity of $v \simeq 51 \pm 14$ km s$^{-1}$ (at 2\,kpc).The magnitude and direction of the proper motion rules out supernova remnant HB9 as the birth-site. We can find no other supernova remnants or groups of massive stars within the region traversed by \sgr\ during its characteristic lifetime ($\sim$20\,kyr).}
   %This is in contrast to other Galactic magnetars which have been traced back to young massive clusters or the centres of supernova remnants.
   {Our observations of \sgr\ suggest that some magnetars may be either significantly older than expected, that their progenitors produce low supernova ejecta masses, or alternatively that they can be formed through accretion-induced collapse (AIC) or low-mass neutron star mergers. Although the progenitor of \sgr\ remains unclear, we propose that \sgr\ is the best Galactic candidate for a magnetar formed through a mechanism other than massive star core-collapse.}
   %{Our observations of \sgr\ suggest that some magnetars may be either significantly older than expected, that their progenitors produce low supernova ejecta masses, or alternatively that they can be formed through the merger or accretion-induced collapse (AIC) of a magnetic white dwarf, or the merger of two low-mass neutron stars. Although the progenitor of \sgr\ remains unclear, we propose that \sgr\ is the best Galactic candidate for a magnetar formed through a mechanism other than massive star core-collapse.}

   \titlerunning{The IR counterpart and proper motion of \sgr}
   \authorrunning{A. A. Chrimes et al.}
   \maketitle

%had to switch to 2020 version of Tex in overleaf to supress bibtex warnings - https://tex.stackexchange.com/questions/625901/overleaf-citation-multiply-defined-with-aa-class-file

%
%-------------------------------------------------------------------
\section{Introduction}
Soft gamma-repeaters (SGRs) are characterized by irregular short bursts of soft $\gamma$- and X-rays, often
repeating on timescales of hours to days, followed by longer periods of inactivity.
Along with Anomalous X-ray Pulsars (AXPs), they are conjectured to be manifestations of magnetars \citep[for reviews see e.g.][]{2008A&ARv..15..225M,2013BrJPh..43..356M,2017ARA&A..55..261K,2021ASSL..461...97E}. 
Magnetars are posited to be young (typical magnetic dipole spin-down ages of $10^3-10^5$\,yr), isolated,
slowly rotating neutron stars (periods typically 2-12~s) with period derivatives in the range $10^{-13}$--$10^{-9}$\,s\,s$^{-1}$ and inferred
magnetic field strengths between $10^{14}-10^{15}$\,G \citep{1998Natur.393..235K,1999ApJ...510L.115K,2014ApJS..212....6O}. There are now around $\sim 30$ examples (confirmed and candidates) within the Milky Way and Magellanic Clouds \citep{2014ApJS..212....6O}\footnote{\url{http://www.physics.mcgill.ca/\~pulsar/magnetar/main.html}}. However, progress in understanding magnetars has been slow due to
the difficulty of observing them in other wavebands and a dearth of
clear evidence for their progenitors and birth places. This is in part a consequence
of their scarcity and locations behind large column densities in the Galactic plane, where most magnetars have been discovered. It also impacts on uncertainties
in their distances \citep[e.g.][]{2006ApJ...650.1070D,2008MNRAS.386L..23B,2020ApJ...905...99Z,2021MNRAS.503.5367B}, and hence birth rates \citep{2019MNRAS.487.1426B}, energetics and kinematics \citep[e.g.][]{2022ApJ...926..121L}.

The known population of magnetars is clearly dwarfed by the young radio pulsar population \citep{2005AJ....129.1993M}. It may be that magnetars are not an uncommon outcome of core-collapse, but that their magnetic field rapidly decays, making them observationally rare \citep{2019MNRAS.487.1426B}. Alternatively, they may be intrinsically rare at birth. Perhaps they come only from more massive progenitors \citep[e.g.][]{2005ApJ...620L..95G}, preferentially arise from progenitors that have undergone a stellar merger \citep[e.g.][]{2020MNRAS.495.2796S,2024MNRAS.531.2379S}, or otherwise have a different, rare progenitor channel \citep[e.g. accretion or merger-induced collapse,][]{1991ApJ...367L..19N,1999ApJ...516..892F,2006MNRAS.368L...1L,2013A&A...558A..39T,2019MNRAS.484..698R,2025ApJ...978L..38C}. Distinguishing these origins based on population statistics alone is challenging due to strong observational biases (e.g., magnetars are typically discovered in outburst).

In principle, studies of magnetar environments and their multi-wavelength spectra can play a major role in understanding their nature. By identifying their natal environments it is possible to independently estimate their age (and progenitor mass) from that of their putative parent stellar population or associated supernova remnant \citep{2005ApJ...622L..49F,2006ApJ...636L..41M,2008MNRAS.386L..23B,2009ApJ...707..844D,2023ApJ...950..137N}, which, in turn, has major implications for the estimate of their birth rates. 

Furthermore, studying the multi-wavelength emission
itself can provide valuable insight into the mechanisms of energy production from the SGR, enabling one to 
discriminate between magnetospheric emission \citep{2007ApJ...657..967B,2011AdSpR..47.1298Z}, and alternatives, including emission from a supernova fallback disc around the magnetar \citep[e.g.][]{2000ApJ...541..344P,2001ApJ...550..397M,2006Natur.440..772W,2009ApJ...700..149K,2016MNRAS.458L.114M,2016ApJ...833..265T,2019ApJ...877..138X,2024ApJ...972..176H} - see also the long-period compact central object 1E\,161348-5055 in supernova remnant RCW\,103 \citep{2006Sci...313..814D}, which may be a magnetar whose rotation period has been slowed to $\sim$6.7\,hours by interaction with a fallback disc \citep{2016ApJ...828L..13R}. A further possibility is a binary companion origin for the optical/near-infrared emission \citep{2008ApJ...681..530P,2016A&AT...29..183P,2022MNRAS.512.6093C,2022MNRAS.513.3550C}.

Magnetars have been well studied at high energies as they are
usually strong persistent X-ray emitters, with X-ray luminosities of
about $10^{34}-10^{36}$\,erg\,s$^{-1}$, thought to be powered by the
ultra-strong magnetic field of these neutron stars \citep{1992ApJ...392L...9D,1993ApJ...408..194T}. They are often detected at hard X-ray and soft gamma-ray energies \citep[not only through SGR behaviour but also via their short/intermediate bursts and giant flares, both Galactic and extragalactic,][]{1979Natur.282..587M,1999AstL...25..635M,2005Natur.434.1107P,2006A&A...449L..31G,2021ApJ...907L..28B,2024Natur.629...58M}, and sometime also at radio frequencies \citep[e.g.][]{2006Natur.442..892C,2007ApJ...669..561C,2007ApJ...662.1198H,2012ApJ...748L...1D,2013MNRAS.435L..29S}.

Despite the observational challenges presented by their dusty, often crowded sight-lines, searches for optical/near-infrared (NIR) counterparts have been partially successful, with a growing sample of candidates being found \citep[e.g.][]{2022MNRAS.512.6093C}. These magnetar counterparts are faint, with magnitudes $K \sim$20--24 \citep{2001ApJ...563L..49H,2002ApJ...580L.143I,2004A&A...425L...5R,2005A&A...438L...1I,2005ApJ...623L.125K,2006ApJ...648..534D,2008A&A...482..607T,2018ApJ...854..161L,2018MNRAS.473.3180T,2022MNRAS.512.6093C}. They can be differentiated from 
confused stellar sources by their unusual colours \citep{2006ApJ...648..534D,2008A&A...482..607T,2022MNRAS.513.3550C}, short time timescale optical variability \citep[at the rotation period,][]{2005MNRAS.363..609D,2009MNRAS.394L.112D,2011MNRAS.416L..16D} and variability on timescales of months to years \citep{2002ApJ...580L.143I,2004A&A...425L...5R,2005ApJ...623L.125K,2006ApJ...652..576D,2022ApJ...926..121L,2022MNRAS.512.6093C}.

Improvements in adaptive optics technology, and the application of the Hubble Space Telescope ({\em HST}) to the problem, have improved the certainty of these identifications by greatly enhancing the spatial resolution of the observations and removing confusion. In turn these IR detections have enabled proper motions to be determined for five magnetars \citep{2012ApJ...761...76T,2013ApJ...772...31T,2022ApJ...926..121L}, demonstrating
velocities of $v_t \sim 100-300$ km s$^{-1}$ (with 4/5 at the lower end of this range). In four out of the five cases these proper motion vectors point back towards either clusters of massive stars (in the case of SGR 1806-20 and SGR 1900+14) or towards 
a recent supernova remnant (in the case of AXP 1E 2259+586 and SGR\,1935+2154). Only in one case \citep[4U\,0142$+$61,][]{2013ApJ...772...31T} does
the proper motion not allow the identification of a likely birth site. A small number of magnetars have a proper motion from very long baseline interferometry (VLBI) radio observations \citep{2007ApJ...662.1198H,2012ApJ...748L...1D,2024ApJ...971L..13D}, but even the highest resolution X-ray observatories such as {\em Chandra} struggle to measure such small proper motions \citep[e.g.][]{2007Ap&SS.308..217M}.  Even when proper motions can be measured, the high column densities and extinctions typical for magnetar sight-lines make associations with supernova remnants or star-forming regions difficult. Nevertheless, these multi-wavelength observations, along 
with the Galactic distribution of magnetar candidates \citep[tightly aligned with the plane][]{2014ApJS..212....6O}, offer strong support for a model in which magnetars are indeed young neutron stars, born in a population of young, and potentially very massive stars. It has been suggested that magnetars could experience large kicks at their formation, caused by the high magnetic fields present in the progenitor \citep{1992ApJ...392L...9D,2008ApJ...672..465S}. In this case we might expect to see rapid proper motion from SGRs and other magnetars, but on the contrary, the growing sample of magnetar velocities shows similar values to young pulsars \citep{2022ApJ...926..121L,2024ApJ...971L..13D}.

Galactic supernova remnants (SNRs) associated with magnetars do not show evidence for additional energy injection \citep[e.g.][]{2006MNRAS.370L..14V,2014MNRAS.444.2910M}, and population synthesis studies disfavour a scenario in which Galactic magnetars produced gamma-ray bursts at the moment of their formation \citep{2015ApJ...813...92R}. This disfavours very short initial rotation periods for Galactic magnetars. One possible explanation for the origin of the strong magnetic fields is a dynamo mechanism \citep[e.g.][]{2008AIPC..983..391S,2022ApJ...926..111W,2022A&A...668A..79B,2024arXiv240701775B,2024arXiv241119328R}. Alternatively, the magnetic field may be inherited from the progenitor \citep[the fossil field scenario, if the progenitor magnetic flux is conserved during collapse,][]{2006MNRAS.367.1323F,2023Sci...381..761S}.

\sgr\, was discovered on 22nd August 2008 by the Neil Gehrels {\em Swift} Observatory (hereafter {\em Swift}) Burst Alert Telescope (BAT), through the
detection of SGR-like bursts \citep{2008ATel.1676....1B,2008GCN..8113....1B,2009MNRAS.396.2419R}. \sgr\ was subsequently observed to exhibit X-ray \citep{2008ATel.1677....1G} and optical pulsations at the rotational period of 5.76s \citep{2011MNRAS.416L..16D}. It has a measured period derivative of $\sim 1.5\times10^{-11}$\,s\,s$^{-1}$, with an inferred magnetic field strength of $\sim 2\times10^{14}$\,G and a characteristic age of $\sim$15\,kyr \citep{2014MNRAS.438.3291C}. The activation of this SGR in 2008 followed a very long period of quiescence \citep{2009MNRAS.396.2419R}. In the days following the onset of activity, tens of bursts were observed, with fluxes exceeding the underlying continuum by a factor $>10^5$. The bursts reached maximum luminosities of $\sim 10^{41}$\,erg\,s$^{-1}$ and had durations of $<$1\,s, typical of short bursts emitted by magnetars \citep[e.g.][]{2008A&ARv..15..225M}. 

In this paper, we report observations of the NIR counterpart to \sgr\ in the months following its initial 2008 outburst, and on the results of a subsequent
$\sim$12 year campaign of NIR monitoring, including a {\em HST}-derived proper motion. The location in the Galactic anti-centre direction, consequent lack of crowding and relatively low extinction make \sgr\ an ideal case for investigating the optical/NIR properties and birth site of a magnetar. Throughout, all magnitudes are reported in the Vega system and all times are UT.

\begin{figure}
\centering
\includegraphics[width=8.5cm,angle=0]{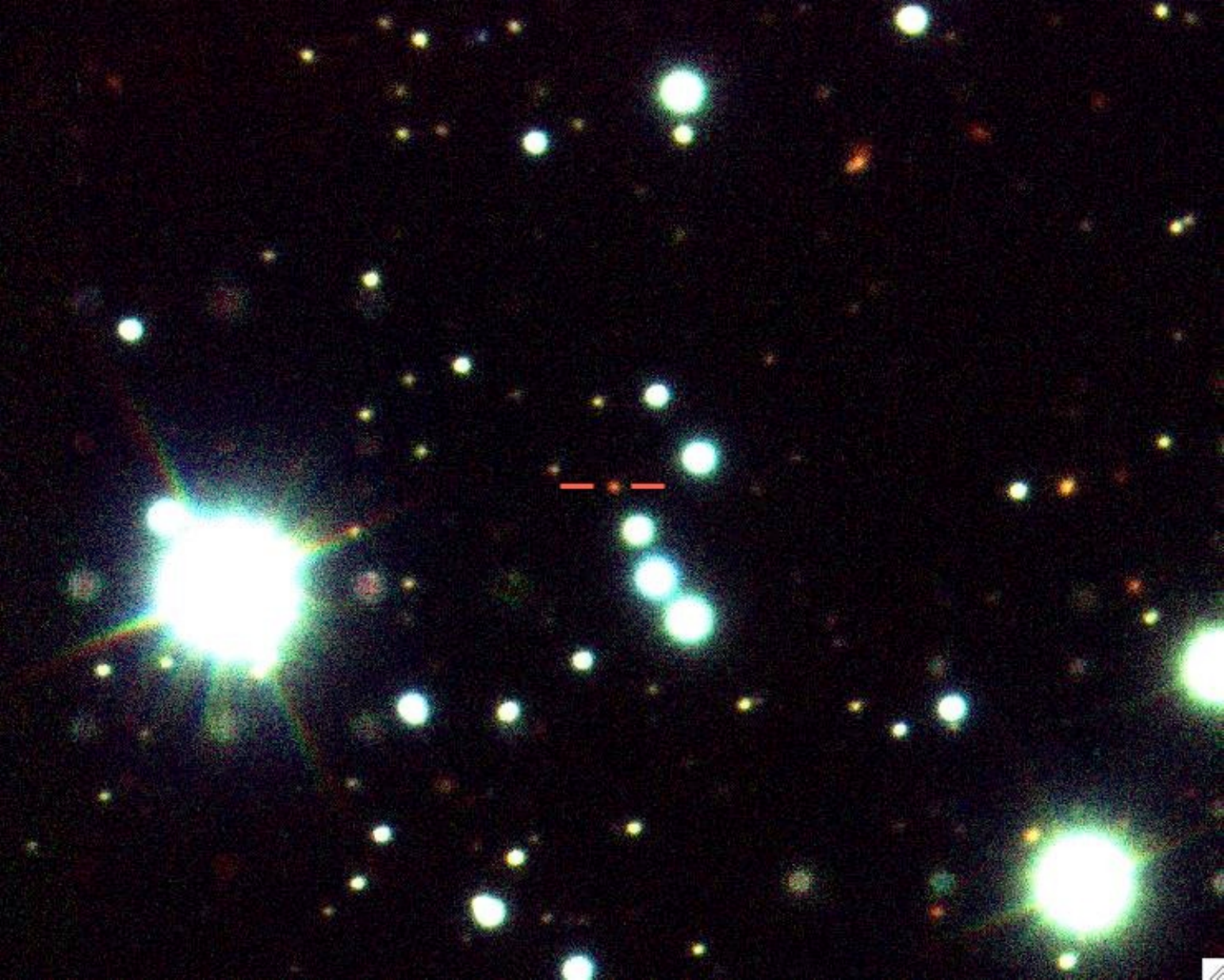}
\caption{\label{figure:finder}A NIR image (JHK-bands) of the field of \sgr\ as imaged with Gemini/NIRI (see Table \ref{table:observations}). \sgr\ is marked with crosshairs and is notably redder than the surrounding field stars. The image has dimensions 45$\times$36\,arcsec and is oriented north up, east left.}  
\end{figure}

\begin{figure*}
\centering
\sidecaption
\includegraphics[width=10cm]{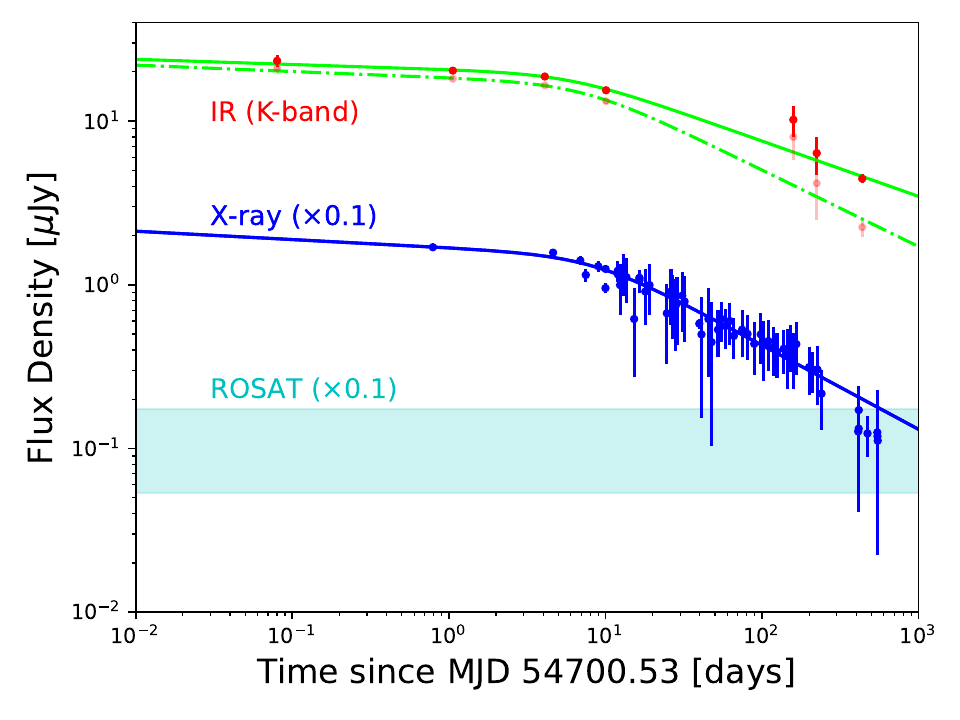}
\caption{The NIR and X-ray lightcurve
of \sgr\ up to $\sim 500$ days post-discovery in 2008. The red points correspond to the NIR and 
the blue to the X-ray from \citet{2009MNRAS.396.2419R}, supplemented with
later time {\em Swift} observations \citep{2018MNRAS.474..961C}. The cyan shaded box 
represents the {\em ROSAT} quiescent level, and its associated error \citep[also from][]{2009MNRAS.396.2419R}. All X-ray fluxes are reduced by a factor of 10 for clarity. The dashed green and solid blue lines are broken power-law fits to the NIR and X-ray data respectively (see Section \ref{sec:lc}). Due to the difficulty of fully accounting
for colour terms between $Ks$ and $K$-band we have plotted only the $K$-band
observations on this figure (red points and solid green line). Also shown, in faded red and with a dashed green line, is the $K$-band light-curve and fit after subtraction of an approximate quiescent flux level as described in the text. The X-ray light-curve is plotted as a specific flux ($F_{\nu}$) at 1\,keV based on the spectral model and count-rate obtained from {\em XMM-Newton} observations over the 0.3-10\,keV range, and assumes no X-ray spectral variability, although such effects have only small effects on the observed flux. \\ \\ \\ }
\label{figure:lightcurve}
\end{figure*}

\section{Observations and analysis}
After the first detection of high energy bursts from \sgr\ by {\em Swift}/BAT,
NIR data were obtained promptly with the 3.8\,m United Kingdom
InfraRed Telescope's (UKIRT) Fast-Track Imager (UFTI) at the Mauna Kea
Observatory, using its rapid-response mode. Subsequent NIR data were obtained over the following 4 years using the
8.1\,m Gemini-North's Near-InfraRed Imager and spectrometer (NIRI) at
Mauna Kea, and the Long-slit Intermediate Resolution Infrared
Spectrograph (LIRIS) on the 4.2\,m William Herschel Telescope at the Roque
de Los Muchachos Observatory. Finally, we obtained three epochs of observations with {\em HST}/WFC3 in October 2010, October 2012 and August 2020, as listed in Table \ref{table:observations}.

Ground based data were reduced using the respective instrument
pipelines. Photometric calibration was performed using the two micron
all-sky survey \citep[2MASS,][]{2006AJ....131.1163S}, which has
also been used for astrometric calibration. The UFTI and NIRI $K$-band
filters are $K$, while the LIRIS and 2MASS $K$-band filters are $K_s$. For UFTI and NIRI photometric calibration, a small correction of 0.02 magnitudes has first been applied to bring
all 2MASS $K_s$-band magnitudes to $K$, using $K_s = K + 0.002 + 0.026 (J -
K)$. The mean $J$-$K$ value for 2MASS field stars within 1\,arcmin is $1.0\pm0.2$ \footnote{Explanatory Supplement to the 2MASS All Sky Data Release
and Extended Mission Products:
\url{http://www.ipac.caltech.edu/2mass/releases/allsky/doc/explsup.html}}. We are limited in the precision of this photometric calibration by uncertainties in the flat fielding corrections, the lack of good NIR flux standards and the scatter in the $K_s$ to $K$ conversion.

We obtained three epochs of observation of SGR\,0501+4516 with {\em HST}/WFC3. The first two were obtained on 19 October 2010 and 8 October 2012, approximately 720 days apart. For these epochs we obtained a single orbit of observation in the F160W (broad $H$-band, effective wavelength 15278\AA) filter. Each observation consisted of four dithered exposures in a standard box pattern, and the orientation was chosen such that diffraction spikes from nearby bright stars did not impinge on the source position. The third epoch was obtained on 4 August 2020, and consisted of F125W (broad $J$-band, effective wavelength 12364\AA) and F160W observations, each with three dithers \citep[for further details, including exposure times, see Table \ref{table:observations} and][]{2022MNRAS.512.6093C}. 

The images were retrieved from the archive\footnote{\url{http://archive.stsci.edu}} and reduced with standard astrodrizzle \citep{2002PASP..114..144F} procedures with {\sc drizzlepac} \footnote{\url{https://www.stsci.edu/scientific-community/software/drizzlepac.html}} using a pixel scale of $0.065"$ per pixel and \texttt{pixfrac} = 0.8. Since we are interested in astrometric fidelity we used the most up to date distortion tables, and determined shifts between each dithered image directly from sources in the image (using tweak-shifts) rather than from the pre-programmed offsets. These images were then drizzled to a common ({\em Gaia} DR3) reference frame and orientation, where the alignment is via cross-correlation of stars with the same 25 {\em Gaia}/DR3 sources at the epoch of each image.

A single source coincident with the position of the X-ray counterpart
to \sgr\ \citep{2008ATel.1691....1W,2014MNRAS.438.3291C,2010ApJ...722..899G} is visible in
all co-added observations. The position of this source, referenced to 2MASS, is
$\rm{RA}\,(2000)=05^{\rm{h}} 01^{\rm{m}} 06\fs75$,
$\rm{Dec.}\,(2000)=45\degr 16\arcmin 34\farcs0$, with an
error of $0.2$ arsecond in both coordinates. This is 0.10 arcseconds
from the centroid of the {\em Chandra} localisation \citep{2010ApJ...722..899G}. We
therefore identify this source as the NIR
counterpart to \sgr; subsequent variability clinches this association. 
For the NIRI data, the source is also detected in
the individual 60 second exposure frames, although at a very low significance level in the
$J$- and $H$-bands. 

Photometry of the ground-based infrared imaging was performed using
IRAF \citep{tody1986:proc:733} aperture photometry routines.  
We calibrated to a sequence of nearby 2MASS stars, ensuring
the accuracy of the relative photometry, our {\em HST} observations
were photometrically calibrated using the published {\em HST} zero points for WFC3/IR.
The results for the SGR counterpart are
provided in Table~\ref{table:observations}.

\begin{figure}
\includegraphics[width=\columnwidth]{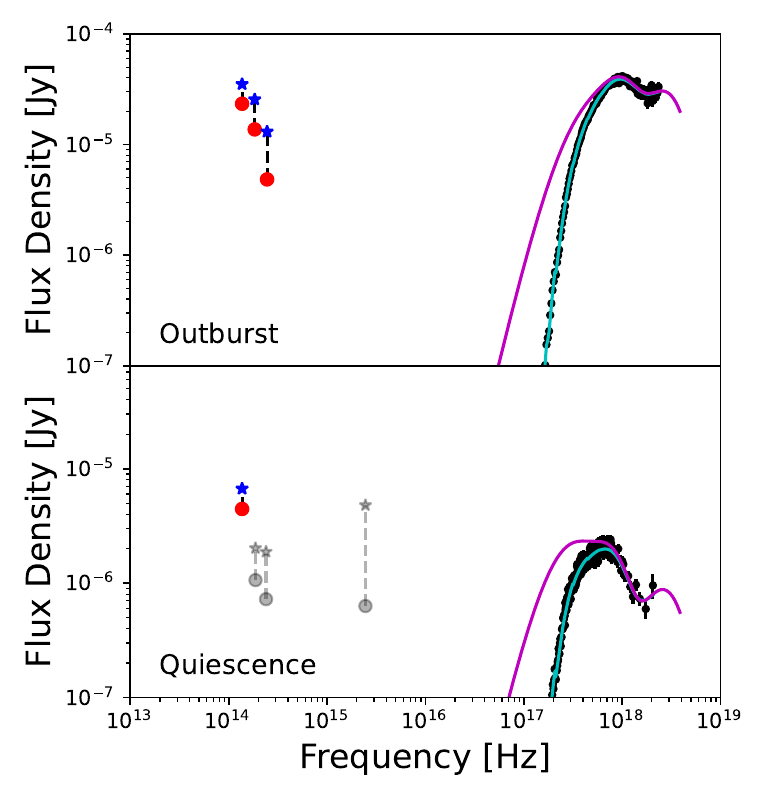}
\caption{\label{figure:sed} Top panel: The spectral energy distribution of \sgr\ from the X-ray to the NIR as measured one day after the first outburst. The X-ray spectra are those obtained by {\em XMM-Newton}, while for the NIR data, the two colours correspond to NIR observations with i) no extinction correction (lower, red points) and ii) extinction corrected (upper, blue points) assuming a total Galactic $E(B-V)=1.3$ in that direction. The true extinction will lie between these extremes (see Figure \ref{fig:ext}).
Bottom panel: The $K$-band and X-ray spectral energy distribution at $\sim 200$ days. Also shown in grey are the late-time $i$-band flux level \citep{2011MNRAS.416L..16D} and the F125W and F160W {\em HST} fluxes from the 2020 epoch. In both panels, the X-ray data are fitted with the models from \citet[][cyan lines]{2009MNRAS.396.2419R}. Also shown (with magenta lines) are unabsorbed (and extrapolated) versions of these models.}  
\end{figure}

\begin{figure}
\includegraphics[width=\columnwidth]{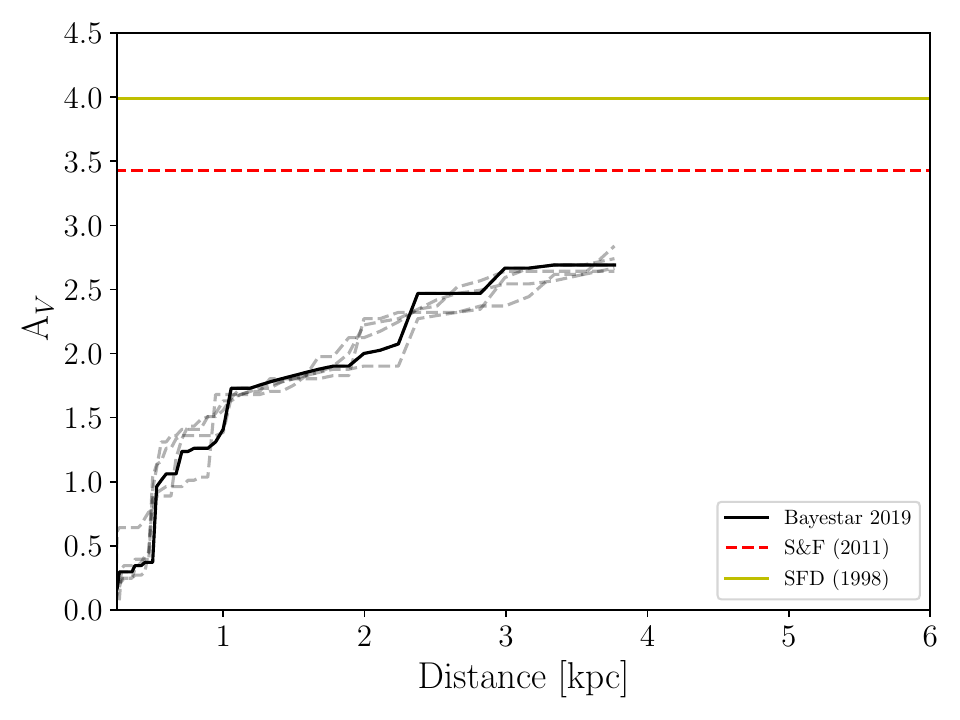}
\caption{\label{fig:ext} Visual extinction $A_V$ as a function of distance along the sight-line to \sgr\,. The solid black line is the best-fit \citet{2019ApJ...887...93G} Bayestar 3D dust map extinction along the sight-line. Dashed grey lines are random draws from the Bayestar probability distribution. Also shown are the total sight-line extinctions from \citet{1998ApJ...500..525S} and \citet{2011ApJ...737..103S}. The Bayestar flattening at $\sim$3.5--4\,kpc corresponds to the point at which main sequence stars can no longer be detected (and the extinction estimate becoming unreliable). The distance to \sgr\ is unknown, but we assume it lies at $\sim$2\,kpc, in the Perseus spiral arm.}  
\end{figure}

%%%%%%%%%%%%%%%%%%%%  TABLE 1  %%%%%%%%%%%%%%%%%%%%%%%%%%%%%%%%%%%%%%%%%%%%%
\begin{table*}
\caption{\label{table:observations}
Differential
photometry of \sgr\ relative to 2MASS stars within the field of view of NIRI, UFTI and LIRIS, and {\em HST} NIR magnitudes for \sgr\ in 2010, 2012 and 2020 (given in the last four rows).
}
\centering
\begin{tabular}{llllcc}
\hline
\hline
Start time & exposure time & filter & magnitude & telescope + instrument\\
Year-month-date-UT &  (seconds) &   & (Vega) &  \\
\hline
%2008 08 / 25\ 03:11:12.0 & 6525 & $K_S$ & $19.12 \pm 0.05$  &  WHT + LIRIS\\
%2008 08 / 31\ 02:31:00.6 & 2085 & $K_S$ & $19.22 \pm 0.11$  &  WHT + LIRIS\\
%2008 09 / 01\ 03:44:35.3 & 2250 & $K_S$ & $19.27 \pm 0.10$  &  WHT + LIRIS\\
%2008 09 / 06\ 04:34:59.6 & 1188 & $K_S$ & $19.05 \pm 0.16$  &  WHT + LIRIS\\
%2008 09 / 20\ 03:43:14.8 & 3735 & $K_S$ & $19.24 \pm 0.15$  &  WHT + LIRIS\\
%2008 08 / 23\ 14:27:03.3 & 1020 & $J$ & $21.02 \pm 0.05$  &  Gemini + NIRI \\
%2008 09 / 01\ 14:28:08.9 & 960  & $J$ & $21.07 \pm 0.07$  &  Gemini + NIRI \\
%2008 08 / 23\ 14:15:25.3 & 480  & $H$ & $19.95 \pm 0.04$  &  Gemini + NIRI \\
%2008 08 / 23\ 14:03:43.8 & 510  & $K$ & $18.78 \pm 0.02$  &  Gemini + NIRI \\
%2008 08 / 26\ 15:04:15.8 & 480  & $K$ & $18.87 \pm 0.01$  &  Gemini + NIRI \\
%2008 09 / 01\ 14:53:07.4 & 900  & $K$ & $19.08 \pm 0.03$  &  Gemini + NIRI \\
2008 08 25\ 03:11:12 & 6525 & $K_S$ & $19.12 \pm 0.05$  &  WHT + LIRIS\\
2008 08 31\ 02:31:01 & 2085 & $K_S$ & $19.22 \pm 0.11$  &  WHT + LIRIS\\
2008 09 01\ 03:44:35 & 2250 & $K_S$ & $19.27 \pm 0.10$  &  WHT + LIRIS\\
2008 09 06\ 04:35:00 & 1188 & $K_S$ & $19.05 \pm 0.16$  &  WHT + LIRIS\\
2008 09 20\ 03:43:15 & 3735 & $K_S$ & $19.24 \pm 0.15$  &  WHT + LIRIS\\
2008 08 23\ 14:27:03 & 1020 & $J$ & $21.02 \pm 0.05$  &  Gemini + NIRI \\
2008 09 01\ 14:28:09 & 960  & $J$ & $21.07 \pm 0.07$  &  Gemini + NIRI \\
2008 08 23\ 14:15:25 & 480  & $H$ & $19.95 \pm 0.04$  &  Gemini + NIRI \\
%2008 08 23\ 14:03:44 & 510  & $K$ & $18.78 \pm 0.02$  &  Gemini + NIRI \\
%2008 08 26\ 15:04:16 & 480  & $K$ & $18.87 \pm 0.01$  &  Gemini + NIRI \\
%2008 09 01\ 14:53:07 & 900  & $K$ & $19.08 \pm 0.03$  &  Gemini + NIRI \\
%2009 01 28\ 07:42:57 & 170 & $K$ & $19.53 \pm  0.27$ & Gemini + NIRI\\
%2009 04 03\ 06:16:53 & 270 & $K$ & $20.04 \pm 0.33$ & Gemini + NIRI \\
%2009 11 02\ 10:57:00 & 3120 &$K$ & $20.43 \pm  0.07$ & Gemini + NIRI \\
%2008 08 22\ 14:31:39 & 270 & $H$ & $19.67 \pm 0.14$  &  UKIRT + UFTI \\
%2008 08 22\ 14:27:52 & 405 & $J$ & $21.28 \pm 0.34$  &  UKIRT + UFTI \\
%2008 08 22\ 14:34:43 & 270 & $K$ & $18.63 \pm 0.09$  &  UKIRT + UFTI \\
2008 08 23\ 14:03:44 & 510  & $K$ & $18.76 \pm 0.07$  &  Gemini + NIRI \\
2008 08 26\ 15:04:16 & 480  & $K$ & $18.85 \pm 0.07$  &  Gemini + NIRI \\
2008 09 01\ 14:53:07 & 900  & $K$ & $19.06 \pm 0.08$  &  Gemini + NIRI \\
2009 01 28\ 07:42:57 & 170 & $K$ & $19.51 \pm  0.28$ & Gemini + NIRI\\
2009 04 03\ 06:16:53 & 270 & $K$ & $20.02 \pm 0.34$ & Gemini + NIRI \\
2009 11 02\ 10:57:00 & 3120 &$K$ & $20.41 \pm  0.10$ & Gemini + NIRI \\
2008 08 22\ 14:31:39 & 270 & $H$ & $19.67 \pm 0.14$  &  UKIRT + UFTI \\
2008 08 22\ 14:27:52 & 405 & $J$ & $21.28 \pm 0.34$  &  UKIRT + UFTI \\
2008 08 22\ 14:34:43 & 270 & $K$ & $18.61 \pm 0.11$  &  UKIRT + UFTI \\
2010 10 19\ 02:37:24 & 2797 & F160W & 22.45 $\pm$ 0.02 & {\em HST} + WFC3/IR \\
2012 10 08\ 19:23:37 & 2797 & F160W & 22.48 $\pm$ 0.02 & {\em HST} + WFC3/IR\\
2020 08 04\ 11:42:00 $\dagger$ & 598 & F125W & 23.33 $\pm$ 0.07 & {\em HST} + WFC3/IR\\
2020 08 04\ 11:55:03 $\dagger$ & 598 & F160W & 22.56 $\pm$ 0.07 & {\em HST} + WFC3/IR\\
\hline
\hline
\end{tabular}
\tablefoot{An additional error of about 0.07 magnitudes has been added in quadrature to the $K$-band magnitude errors: this is the scatter in the transformation of 2MASS reference star magnitudes from $K_s$ to $K$, using $K_s = K + 0.002 + 0.026 (J-K)$. Magnitudes are not corrected for Galactic extinction (see Fig.~\ref{fig:ext}). $\dagger$ - Photometry from \citet{2022MNRAS.512.6093C}.}
\end{table*}

%%%%%%%%%%%%%%%%%%%%%%%%%%%%%%%%%%%%%%%%%%%%%%%%%%%%%%%%%%%%%%%%%

We additionally searched for short time scale variability, on the time scales
permitted by the individual frames (NIRI; individual frames with
exposure times of 30 -- 60 seconds) or sub-coadded frames (UFTI;
integration times of 5 minutes). We find no evidence for such
variability: the source remains essentially constant over periods of
15 -- 30 minutes. However, much higher cadence observations ($<1$ second) 
in both the optical and infrared have revealed coherent pulsations with a 
period of 5.7 seconds, identical to the X-ray derived period \citep{2011MNRAS.416L..16D}, thus confirming the counterpart identification.

\section{Properties of the NIR counterpart of SGR0501+4516} 
\subsection{The infrared light-curve}\label{sec:lc}
The infrared light-curve of \sgr\ is shown in Figure~\ref{figure:lightcurve}, in addition to the X-ray
light-curve from \citet{2009MNRAS.396.2419R} for comparison. The X-ray light-curve is extrapolated to a specific flux at
1\,keV based on the power-law plus black body model obtained for the outburst phase by \citet{2009MNRAS.396.2419R}. T$_{0}$ is defined as August 22.53 2008 (MJD\,54700.53), which is the time of the {\em Swift}/BAT trigger. The IR light-curve shows a prolonged plateau, with a slow fading, followed by a more rapid fading. This can be fit with a smoothly broken power-law, where $\alpha_1$ and $\alpha_2$ are the pre and post-break decay slopes defined as $F_{\nu} \propto t^{-\alpha}$ and $t_b$ is the break time. We find $t_b = 6.1\pm2.6$ days, with decay indices of $\alpha_1 = 0.03 \pm 0.03$ and $\alpha_2 = 0.34 \pm 0.03$. This can be compared
to the behaviour of the X-ray flux over the same temporal baseline, which yields $t_{b,x} = 8.3\pm0.3$, with slopes of $\alpha_1 = 0.05 \pm 0.05$ and $\alpha_2 = 0.51 \pm 0.03$, suggesting that
the X-ray and IR light curves are broadly tracking each other, although the IR decay is
slower than the X-ray post-break. This may suggest evolution in the black-body$+$power-law spectrum observed in outburst \citep{2009MNRAS.396.2419R}. At very late times, the IR lightcurve appears to plateau at around m(F160W)$\sim$22.5 (see Table \ref{table:observations}), suggesting that these observations have reached a quiescent level. Since these are F160W ($H$-band) observations rather than $K$, we cannot directly subtract the flux. However, assuming a power-law in f$_{\nu}$ determined by F160W-F125W, we can estimate the late-time $K$-band flux as approximately $\sim$2.2$\mu$Jy. Adopting this as the quiescent $K$-band contribution at all times pushes the IR decay closer to the X-ray decay rate, with fit parameters $t_b = 7.4\pm3.4$ days, $\alpha_1 = 0.04 \pm 0.04$ and $\alpha_2 = 0.47 \pm 0.07$.

\subsection{Spectral energy distribution}
The spectral energy distribution of \sgr\ one day after outburst is shown in Figure~\ref{figure:sed}. 
The {\em XMM-Newton} spectra are those obtained from early and late observations, 
and have been reduced as described in \citet{2009MNRAS.396.2419R}. 
The figure shows the NIR fluxes as observed, and 
as they would appear corrected
for a Galactic extinction of $A_{\rm V}=4$ with a \citet{1999PASP..111...63F} extinction law and R$_{\rm V}=3.1$. However, since \sgr\ lies within the Galactic disc it is not necessarily appropriate to correct
the observed fluxes for the (highest estimate) of the total Galactic column from \citet{1998ApJ...500..525S}, and the true fluxes will lie
between the two values, as demonstrated in Figure \ref{fig:ext}. 

Another estimate for the line-of-sight reddening is based on the X-ray
hydrogen column density, which was found to be $(0.89\pm0.01) \times 10^{22}$ cm$^{-2}$ \citep{2009MNRAS.396.2419R}. We can convert this to an extinction using A$_{V}$-N$_{H}$ relations. For instance the relation of \citet{1995A&A...293..889P} yields A$_{V}\sim4.9$ for \sgr\, while \citet{2008AIPC.1000..200S} gives $A_V \sim 4.7$. These values compare with total extinction values of $\sim$3.4 and $\sim$4 from the \citet{2011ApJ...737..103S} and \citet{1998ApJ...500..525S} Galactic dustmaps, respectively. The Bayestar 2019 3D dustmap \citep{2019ApJ...887...93G} yields A$_{V} \sim 2.7$ at $\sim$2\,kpc, the approximate distance of the Perseus arm on this line of sight. The high $A_V$ inferred from the X-ray may suggest a relatively large distance for the SGR (i.e. looking through much of the dust and gas in the direction), but the inferred extinctions are in excess of even the total sight-line extinction, so may instead indicate a high N$_{H}$ column local to the X-ray emitting region.

The counterpart has markedly non-stellar
colours (Figure \ref{figure:finder} and Table \ref{table:observations}), and lies well away from the stellar locus \citep[in common with most other magnetar NIR counterparts, see e.g.][]{2022MNRAS.513.3550C}. Between the two Gemini/NIRI epochs on 23 August 2008 and 1 September 2008, the counterpart
(in addition to fading) showed evidence for becoming bluer in the infra-red bands, specifically evolving from $J-K=2.24\pm0.05$
to $J-K=1.99\pm0.08$.

\section{Proper motion measurement}\label{sec:pm}
\begin{figure*}
\centering
\includegraphics[width=\textwidth]{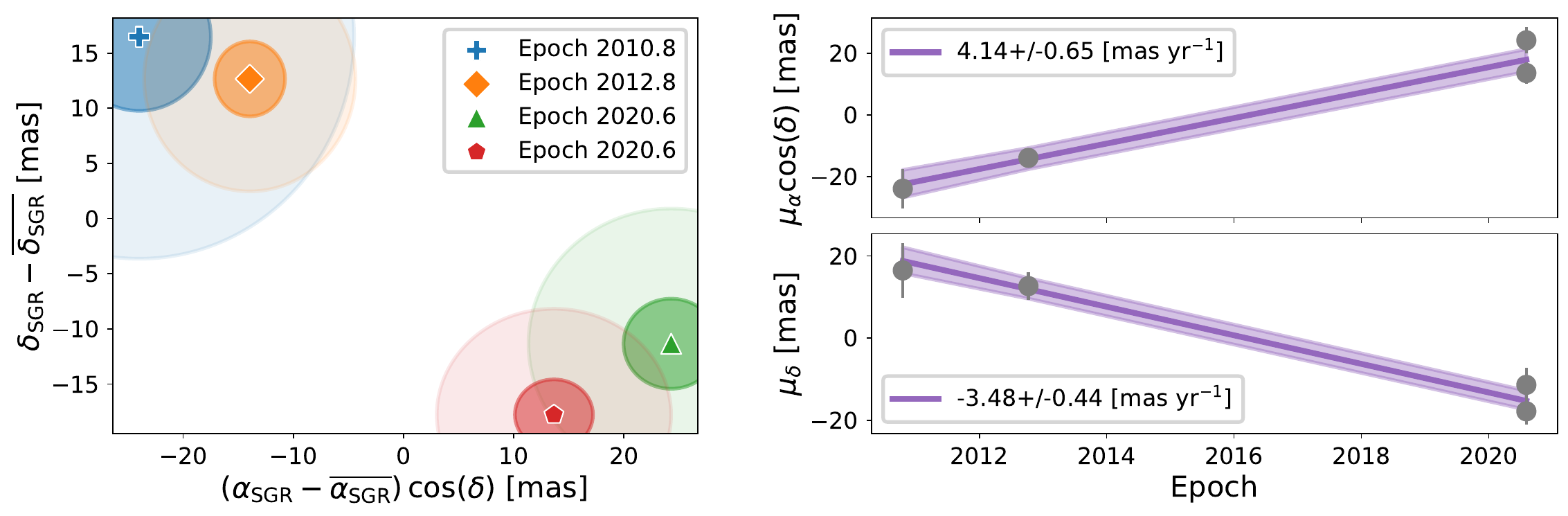}
\caption{\label{figure:pm}Left: offsets of \sgr\ in 2D-projected equatorial coordinates, in each of the three {\em HST} epochs (note that epoch 2020.6 has observations two filters), with respect to the mean position across all four sets of observations. Dark and light shaded ellipses correspond to the 1 and 3\,$\sigma$ uncertainties respectively. Right: linear fits to positional offsets of \sgr\ in $\alpha_{\star}$ and $\delta$, including corrections for Solar and Galactic motion, as a function of time. The uncertainties on the points show the overall positional uncertainties on \sgr\ at each epoch, including residuals from the astrometric tie to the {\em Gaia} EDR3 reference frame. The two 2020 points are from the two filters, and the right-most marker in the right-hand panels has been arbitrarily shifted by 0.15 years for visual clarity. The best-fit proper motions in each coordinate are indicated, and a shaded region indicated the 1$\sigma$ uncertainty on the fit.} 
\end{figure*}

\begin{figure*}
%\centering
\sidecaption
\includegraphics[width=10cm]{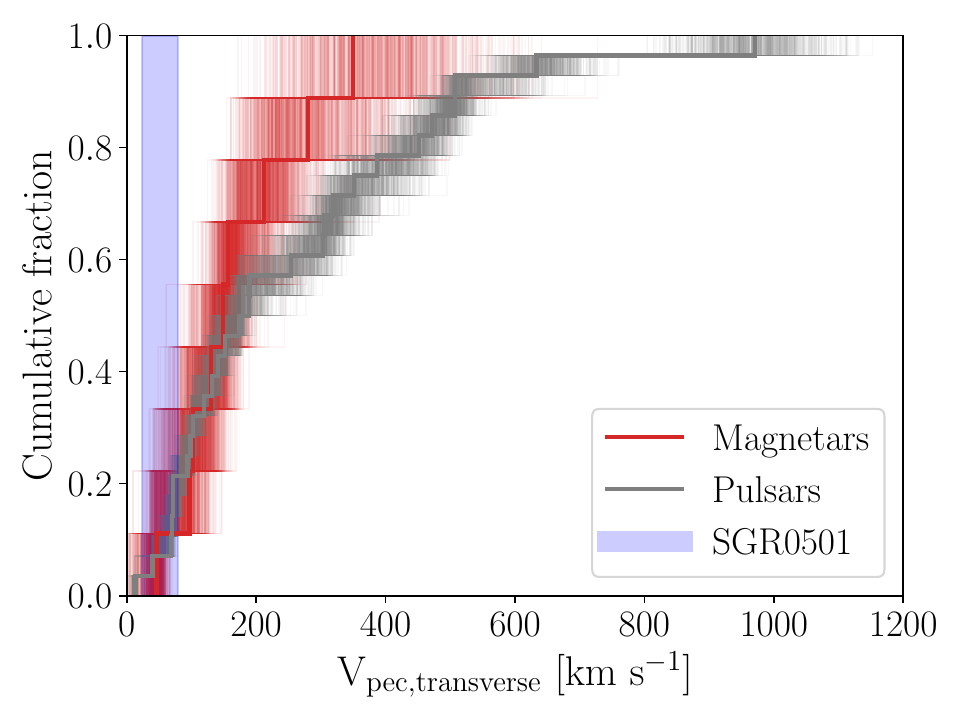}
\caption{The cumulative distribution of the nine magnetar peculiar transverse velocities, including \sgr, measured so far. These measurements are for XTE\,J1810-197 \citep[][VLBI]{2007ApJ...662.1198H}, 1E\,1547.0-5408 \citep[][VLBI]{2012ApJ...748L...1D}, SGRs\,1806-20 and 1900+15 \citep[][NIR]{2012ApJ...761...76T}, AXPs 1E\,2259+586 and 4U0142+61 \citep[][NIR]{2013ApJ...772...31T}, SGR\,1935+2154 \citep[][NIR]{2022ApJ...926..121L}, Swift\,J1818.0-1607 \citep[][VLBI]{2024ApJ...971L..13D}, and \sgr\ (this work). Multiple draws from each magnetar's velocity probability distribution are made, assuming Gaussian uncertainties on the velocities. The same process is applied to the pulsar distribution in grey \citep[data from][]{2017A&A...608A..57V}. \sgr, whose velocity range for distances between 1--3\,kpc is indicated by the shaded band, is the joint-slowest magnetar yet discovered, at least in projection, and is among the slowest $\sim$10\% of pulsars with similar measurements. \\ \\ \\ }
\label{figure:vt}
\end{figure*}

Measuring proper motions for magnetars can constrain their space velocities
and possible birth sites, both of importance for understanding their origins. Well spaced X-ray observations
with {\em Chandra} \citep[e.g. SGR\,1900+14 and AXP\,1E\,2259+586,][]{2009ApJ...692..158D,2009AJ....137..354K} have only yielded upper
limits for magnetars, although X-ray proper motions are possible in principle if the motion is substantial enough \citep[e.g.][]{2024ApJ...976..228R}. There are only a few measurements arising from very long-baseline
radio interferometry \citep[AXP\,XTE\,J1810-197, 1E\,1547.0-5408 and Swift\,J1818.0-1607;][]{2007ApJ...662.1198H, 2012ApJ...748L...1D,2024ApJ...971L..13D}. Ground-based NIR adaptive
optics imaging of SGRs\,1806$-$20 and 1900$+$14 have provided direct measurements of
their proper motions, strengthening suggestions of their origin in young stellar clusters \citep{2012ApJ...761...76T}. 
Similar velocity constraints have also been placed on AXP\,1E2259$+$586 and AXP\,4U0142$+$61 \citep{2013ApJ...772...31T}. In
the case of AXP\,1E2259$+$586, the proper motion makes a compelling case for an origin in the supernova remnant CTB\,109, although in the latter example of AXP\,4U0142$+$61 it has not been possible to find
an association with either a young star cluster or SNR. Since then, the identification of the NIR counterpart of SGR\,1935$+$2154 with {\em HST} \citep{2018ApJ...854..161L} and subsequent epochs of observations have enabled a NIR proper motion measurement for this object too \citep{2022ApJ...926..121L}. Tracing back the proper motion over the characteristic lifetime of SGR\,1935$+$2154 places it at the centre of SNR\,G57.2$+$08.
%Eight in total as of 2024, 3 VLBI and 5 from NIR counterpart. SGR0501 is the ninth

SGR\,0501$+$4516 offers a new opportunity to obtain a magnetar proper motion measurement. To determine its proper motion, we use the method of \citet{2022ApJ...926..121L} and the three epochs of observations with {\em HST}, spread across a temporal baseline of ten years (2010--2020, see Table \ref{table:observations}). The source is well detected in all three epochs with broadly consistent magnitudes of m(F160W) = 22.45 $\pm$ 0.02, 22.48 $\pm$ 0.02 and 22.56 $\pm$ 0.07 (Vega system, statistical uncertainties only). This corresponds to signal-to-noise ratios (SNRs) of 78, 68 and 26. The first two observations have longer exposure times and yield higher SNRs, allowing the centroid of the source \citep[measured with DOLPHOT,][]{2000PASP..112.1383D} in a single image to be determined to better than 0.05 pixels (FWHM / 2.3 $\times$ S/N) or $<1$ mas (statistical uncertainty only, per axis). 

For a proper motion measurement we must define an astrometric reference frame. It is standard to do this in a relative sense, with the use of common tie points in each image, and indeed the moderate Galactic latitude of \sgr\ does allow for sufficient stars to be used as such. However, utilizing this approach with {\em HST} will produce an alignment that is limited by the genuine proper motions of the stars and other effects beside the limitations of the data quality (which is dictated primarily by instrument stability and capability to accurately centroid sources). Instead, we follow the method presented in \citet{2022ApJ...926..121L}, itself building upon \citet{bf18}. Succinctly, each image is aligned to the {\em Gaia} absolute astrometric frame \citep{gaiaedr3}, using the positions and proper motions of stars to determine an epoch-corrected equatorial solution. After fitting for this alignment solution \citep[for details, we refer the reader to][]{2022ApJ...926..121L}, the positions of stars could be freely converted between coordinate systems, including onto each {\em HST} image's ($X$, $Y$) pixel coordinate plane. 

Two residual effects in the position of stars remain at this point owing to Solar motion and Galactic rotation. A simple model of the Galaxy's rotation, including the Sun's position and peculiar motion is used to remove this effect \citep[see][]{2022ApJ...926..121L}. Finally, stars are robustly cross-matched between epochs, with the RMS residual on these cross-matches giving our alignment uncertainty. The {\em Gaia}-{\em HST} offset uncertainties (68$^{\rm th}$ percentiles) are 9.2, 3.7, 5.7 and 3.1\,mas for the four epochs in chronological order. The total uncertainty on the source position in a given epoch is therefore given by the quadrature sum of the centroid and astrometric alignment uncertainties. The shifting coordinates of \sgr\ within the absolute frame set by {\em Gaia} then provides our proper motion constraints. The results of this analysis are shown in Figure \ref{figure:pm}.

Our results show that, after correction for Galactic and Solar motion, \sgr\ has a transverse peculiar motion (i.e. with respect to its local standard of rest) of \mbox{$\mu_\alpha \cos(\delta) = 4.14\pm0.65$} and \mbox{$\mu_\delta = -3.48\pm0.44$\,mas\,yr$^{-1}$}. At a distance of 2\,kpc, this equates to a tangential velocity of \mbox{$51\pm14$\,km\,s$^{-1}$}, although the somewhat unconstrained distance to the source also adds a significant source of additional uncertainty. 
We show this in Figure \ref{figure:vt} for a reasonable distance bounds, and how the inferred velocity compares with the broader populations of magnetars and pulsars. \sgr's tangential velocity of \mbox{$51\pm14$\,km\,s$^{-1}$} is a low value for pulsars and similar to the transverse velocity of magnetar Swift\,J1818.0-1607 \citep{2024ApJ...971L..13D}, which is convincingly associated with a radio-detected SNR based on a common distance and Swift\,J1818.0-1607's past trajectory \citep{2023ApJ...943...20I,2024ApJ...971L..13D}. The peculiar transverse velocity of \sgr\, is among the lowest $\sim 10\%$ of pulsars \citep[e.g.]{2005MNRAS.360..974H}, although we cannot rule out a larger radial component.

%%%%%%%%%%%%%%%%%%%%  Figure %%%%%%%%%%%%%%%%%%%%%%%%%%%%%%%%%%%%%%%%%%%%%

\begin{figure*}
\centering
\includegraphics[width=0.95\textwidth]{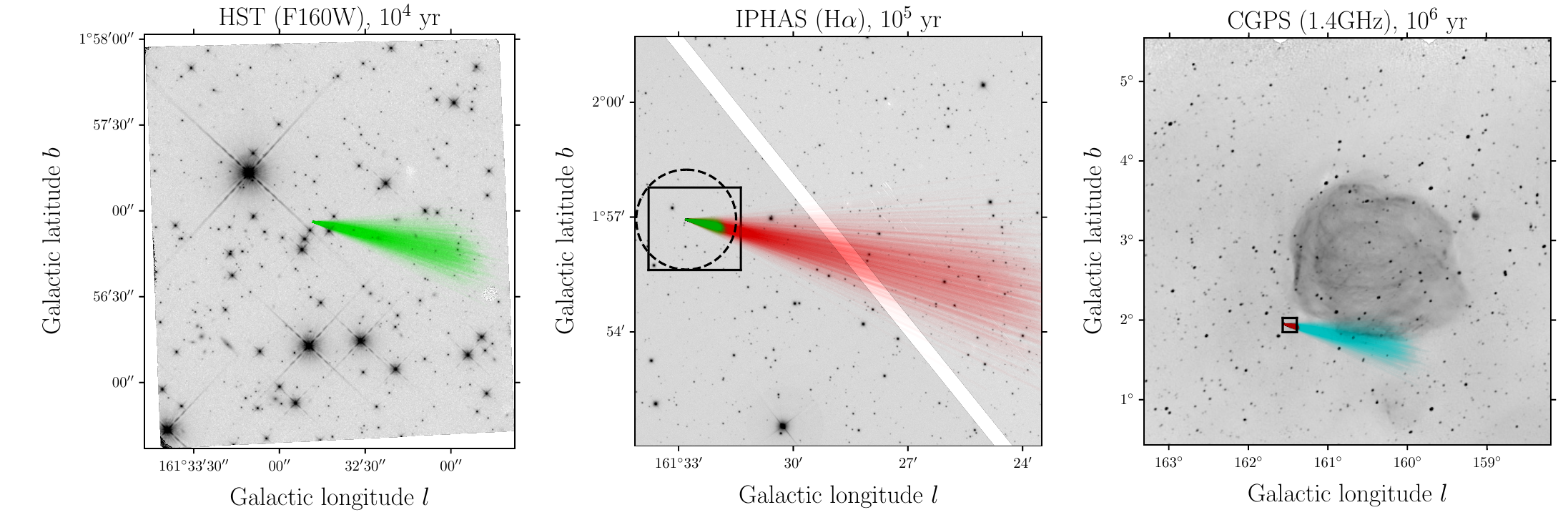}
\caption{\label{figure:location} The possible range of past trajectories of SGR\,0501+4516, based on the proper motion as measured with {\em HST}. The left hand panel shows a {\em HST}/WFC3 F160W image
of the field surrounding \sgr. The possible past trajectories of \sgr\ are indicated by the light green transparent lines. These are random draws from the RA and Dec proper motion distribution, defined by their best-fit values and uncertainties (assuming these are Gaussian), multiplied by an age of $\tau \sim 10$\,kyr. The middle panel shows the same, for wider field of view and IPHAS H$\alpha$ imaging. The past trajectory over $10^{5}$ years is shown in red, the trajectory of $10^{4}$ years - as in the left panel - is shown in green. A dashed circle with radius 15\,kyr $\times$ 5.4\,mas\,yr$^{-1}$ is also shown. The right-hand panel shows a wide-field 1.4\,GHz image from the Canadian Galactic Plane Survey. The trajectory over 1Myr is shown in light blue ($10^{5}$ again in red), demonstrating that \sgr\ and SNR HB9 are not physically associated. In each case, the field of view of the previous panel is shown. As discussed in the text, there is no evidence for any region of intense
star formation within this area which might have given rise to the progenitor star of \sgr. The dark points on the radio image are compact sources, the vast majority of these are expected to be extragalactic \citep{1996A&AS..115..345L}.}
%For scale, the largest of the concentric circles are 40 arcmin in radius.}  
\end{figure*}

%%%%%%%%%%%%%%%%%%%%%%%%%%%%%%%%%%%%%%%%%%%%%%%%%%%%%%%%%%%%%%%%%%%%%%%%%%%%

\section{The birthplace of SGR0501+4516}\label{sec:birthsite}
In Figure \ref{figure:location} we show the extrapolated
vector of the SGR proper motion, and its plausible birth sites for ages of 10\,kyr, 50\,kyr and 1\,Myr. In each case we have
examined broadband optical/IR imaging \citep[from our {\em HST} imaging, DSS and the WFCAM Galactic Plane Survey,][]{2008MNRAS.391..136L}, H$\alpha$ imaging from the INT/WFC Photometric H$\alpha$ Survey of the Northern Galactic Plane \citep[IPHAS,][]{2005MNRAS.362..753D,2021A&A...655A..49G} and 1.4 GHz \citep[Canadian Galactic Plane Survey,][]{2003AJ....125.3145T} radio observations \citep[unfortunately SGR\,0501 lies outside the MeerKAT Galactic plane survey footprint,][]{2025A&A...693A.247A}. It is important to note that age estimates for \sgr\, \citep[$\sim$10--20\,kyr,][]{2014MNRAS.438.3291C} are already at the lower-end of the $10^{4}$ to $10^{6}$\,yr range shown in Figure \ref{figure:location}. This means that the region containing the likely birth-site is commensurately smaller.

\subsection{Association with supernova remnant HB9}
The position of SGR\,0501+4516 is close to the Galactic SNR HB9, whose centre lies roughly 80\arcmin~ from
the SGR. \citet{2008GCN..8149....1G} suggested that the two could both be remnants of the same progenitor. It is then reasonable to ask what is the probability of finding \sgr\, so close (in projection) to a known SNR by chance. Taking the Galactic SNR catalogue of \citet{2019JApA...40...36G}, we select all SNRs within $\pm$5\,deg of the Galactic plane and $\pm45$\,deg of the Galactic anti-centre in longitude. There are 13 SNRs in this portion of the sky, or $\sim$0.01 SNRs\,deg$^{-2}$. Taking the offset of \sgr\ to HB9, we can use a probability of chance alignment argument, where P$_{\rm chance} = 1 - e^{\Sigma \pi r^{2}}$, $\Sigma$ = SNRs\,deg$^{-2}$ and $r$ is the offset in degrees \citep{2002AJ....123.1111B}. In this case, P$_{\rm chance}\sim0.05$, so the simple proximity of \sgr\ and HB9 on the sky does not offer strong support for their association.

HB9 lies at an estimated distance of $\sim 1$ kpc, and a suggested age of $t_{\rm SNR} = 4000-7000$ years \citep{2007A&A...461.1013L}. This age is somewhat younger than the characteristic age of \sgr. While uncertainties in both ages might suggest that such an association should be considered, our measured proper motion vector rules out an association with HB9 at high significance. It lies in the wrong direction, and even if it did not, with a projected velocity of $\sim$50\,km\,s$^{-1}$, the magnetar could only reach such a large projected angular offset in a lifetime of $\sim 10^6$ years. This is much longer than the plausible age of the SN remnant and of the estimated age of \sgr. For \sgr\ to reach an 80\,arcmin offset within the age of the SNR, it would need a kick velocity of several thousands of km\,s$^{-1}$.

A physical association between \sgr\ and HB9 is therefore ruled out by both the magnitude and direction of the proper motion. It is interesting to note that HB9 appears to contain a different neutron star (pulsar PSR\,B0458$+$46). This might point to a core-collapse origin for that SNR, and hence recent star-formation nearby \sgr.

However, there are two arguments for pulsar PSR\,B0458$+$46 also being physically unassociated with HB9. First, its inferred age is much greater than that of the HB9 SNR \citep{2007A&A...461.1013L}; second, distance estimates for HB9 place it at $\sim$1\,kpc (in or in front of the Perseus arm), while the latest distance estimates for PSR\,B0458$+$46 place it well behind the Perseus arm \citep[$>2.7$\,kpc,][]{2023MNRAS.523.4949J}. Therefore, HB9 and PSR\,B0458$+$46 may only be co-located in projection. In the absence of any other candidate neutron star in HB9 and the lack of potential massive star birth sites nearby (see the remainder of this Section), it is possible that HB9 is a supernova Ia remnant, and the tension between \sgr\ as a product of massive star core-collapse and its environment remains.

\subsection{Stellar clusters and stars with overlapping past trajectories} 
The positional uncertainty of \sgr\ at any time $t$ in the past is roughly an ellipse, whose ellipticity and orientation is set by the uncertainties on the position and proper motion, and whose size increases linearly with time. In the left-hand panel of Figure \ref{figure:location}, there are numerous stars in the field of view, but relatively few along the possible past trajectory of \sgr, and none of these appear to reside in obvious bright clusters. Given the excellent PSF of our {\em HST} observations ($\sim 0.1$\arcsec) it is unlikely that these stars are masking a compact stellar cluster, which we would
expect to observe. 

At a distance of 1.5-4\,kpc and extinctions up to $A_{V}\sim3$, O or B stars would have NIR apparent magnitudes $\lesssim 16$ \citep{2022MNRAS.513.3550C}, so our {\em HST} observations should therefore be able to detect any young star-forming regions or clusters, but none appear to be present. The characteristic age and proper motion of \sgr\ mean that a host cluster should be visible and in close proximity (if present). Indeed, there are no O-stars (with $T_{\rm eff} \gtrsim 30000$\,K) within a few degrees of \sgr\ in the StarHorse {\em Gaia} catalogue of \citet{2019A&A...628A..94A}. Young clusters have been suggested as the birthplaces of SGR\,0526-66, SGR\,1900+14, SGR\,1806-20 and CXOU\,J1647-45
\citep{2004ApJ...609L..13K,2000ApJ...533L..17V,2008MNRAS.386L..23B,2014A&A...565A..90C}, but, if the age estimate is broadly correct (it is likely to be an upper limit, see Section \ref{sec:oldmagnetar}), we can confidently rule out such a scenario for \sgr.

\begin{figure*}
\centering
\includegraphics[height=6.5cm,angle=0]{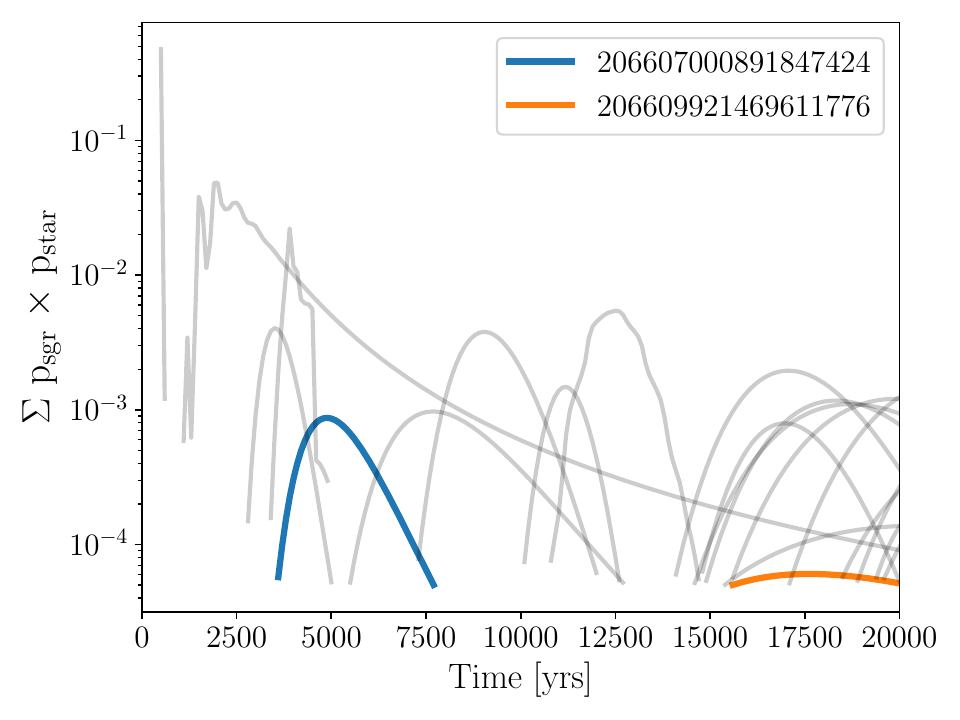}
\includegraphics[height=6.5cm,angle=0]{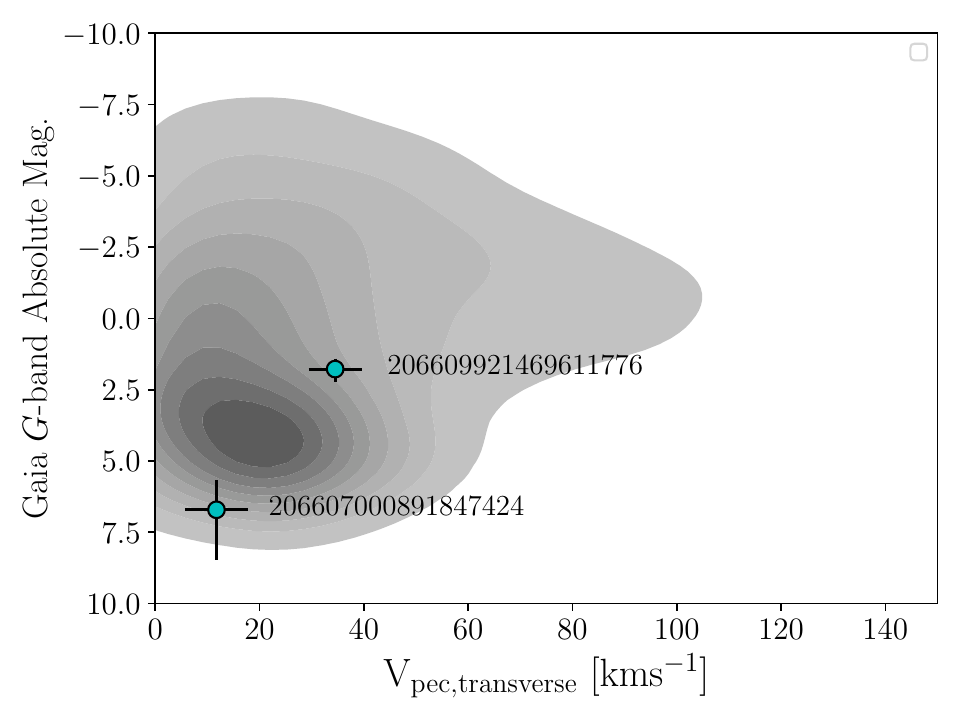}
\caption{\label{figure:comp} Left: the product of overlapping, elliptical 2D Gaussians representing the positional uncertainty of \sgr\, and {\em Gaia} sources in the vicinity, as a function of time. Only sources with 1D, radial separations with more than 3$\sigma$ significance are shown. We find two {\em Gaia} objects that meet this criteria (orange and blue). The grey lines are {\em Gaia} matches found when the SGR proper motion vector is rotated by 90, 180 and 270 degrees. The `real' matches are not obviously distinct from the false/random matches. Right: the two `real' matches, with their extinction-corrected absolute magnitudes (using their best parallax distances and extinction at that distance in the Bayestar 2019 dust map) plotted against their peculiar transverse velocity. Their {\em Gaia} DR3 IDs are labelled. Binary population synthesis predictions for the properties of stars ejected by the supernova of a binary companion are shown by the grey shading \citep{2023MNRAS.522.2029C}, where each contour moving from dark to light grey bounds 10\% more of the probability.}  
\end{figure*}

In case of CXOU\,J1647-45, located in the cluster Westerlund 1, there is also a runaway star \citep[space velocity $>30$\,km\,s$^{-1}$,][]{1961BAN....15..265B} which has been linked with the magnetar progenitor system. Runaway stars can be produced either dynamically \citep[i.e., they are ejected from a cluster through dynamical interactions,][]{1967BOTT....4...86P,1986ApJS...61..419G} or through ejection by the supernova of a binary companion \citep[e.g.][]{2011MNRAS.414.3501E,2014ApJS..215...15S,2019A&A...624A..66R,2023MNRAS.522.2029C}. We therefore searched for stars in {\em Gaia} whose past trajectories intersect the past trajectory of \sgr\, within the last 20\,kyr \citep[age estimates for \sgr\ are in the range 10--20\,kyr,][]{2014MNRAS.438.3291C}. For this, we made use of {\em Gaia} data release 3 \citep{2023A&A...674A...1G}, evolving the positional uncertainty ellipses of stars within 0.5\,deg of \sgr\ back in 100 equal time-steps over 20\,kyr. We only consider sources with a proper motion and parallax measurement, and whose distance lies in the range $1<{\rm d/kpc}<3$.

We determined if \sgr\ is spatially consistent with each {\em Gaia} star at each time-step by calculating the value of a 2D, elliptical Gaussian distribution, corresponding to the probability of finding the object at that position, at each position across the 0.5\,deg $\times$ 0.5\,deg area considered. The ellipsoidal region of positional uncertainty is rotated with respect to the projected RA $\alpha_{\star}$ and Dec $\delta$ axes (where $\alpha_{\star} = \alpha \times {\rm cos}(\delta)$) according to the {\em Gaia} {\sc pmra\_pmdec\_corr} parameter, where the rotation is defined as $\theta_{\rm c}={\rm cos}^{-1}$({\sc pmra\_pmdec\_corr}). The probability density function for a source with best-fit coordinates ($\overline{\alpha_{\star}},\overline{\delta}$) and uncertainties ($\sigma_{\alpha_{\star}},\sigma_{\delta}$) at projected equatorial coordinates ($\alpha_{\star}$, $\delta$) is then given by,
\begin{multline}
    p_{\rm star}(\alpha_{\star},{\delta}) = (2 \pi \sigma_{\alpha_{\star}} \sigma_{\delta})^{-1} \\
    \times {\rm exp}( - (A(\alpha_{\star}-\overline{\alpha_{\star}})^{2} + B(\alpha_{\star}-\overline{\alpha_{\star}})(\delta-\overline{\delta}) + C(\delta-\overline{\delta})^{2}) ),
\end{multline}
\noindent where
\begin{multline}
    A = \frac{{\rm cos}^{2}({\theta_{\rm c}})}{2\sigma_{\alpha_{\star}}^{2}} + \frac{{\rm sin}^{2}({\theta_{\rm c}})}{2\sigma_{\delta}^{2}}, \\
    B = \frac{-{\rm sin}^{2}(2{\theta_{\rm c}})}{4\sigma_{\alpha_{\star}}^{2}} + \frac{{\rm sin}^{2}(2{\theta_{\rm c}})}{4\sigma_{\delta}^{2}}, \\
    C = \frac{{\rm sin}^{2}({\theta_{\rm c}})}{2\sigma_{\alpha_{\star}}^{2}} + \frac{{\rm cos}^{2}({\theta_{\rm c}})}{2\sigma_{\delta}^{2}}. \\
\end{multline} %https://en.wikipedia.org/wiki/Gaussian_function#Two-dimensional_Gaussian_function

\noindent The product of each star's and \sgr\,'s probability density function is calculated, and then integrated
across the whole area, such that the total probability $P$ of \sgr\ and any given {\em Gaia} star being spatially coincident at each time-step is given by,
\begin{multline}
    P = \int_{\alpha_{\star,min}}^{\alpha_{\star,max}} \int_{\delta_{min}}^{\delta_{max}} p_{\rm star}(\alpha_{\star},{\delta}) \times p_{\rm sgr}(\alpha_{\star},{\delta})  d{\delta} d{\alpha_{\star}}.
\end{multline}
In Figure \ref{figure:comp} we show the two {\em Gaia} sources whose separations from \sgr\, at some point in the last 20\,kyr yield a probability of being spatially coincident of $10^{-5}$ or greater. The probability of these objects being associated with \sgr\ never exceeds $\sim 10^{-3}$. The peculiar transverse velocities (calculated with {\em Gaia} distances from the parallax and proper motions) and $G$-band magnitudes are shown in the right-hand panel of Figure \ref{figure:comp}, overlaid on Binary Population and Spectral Synthesis \citep[BPASS,][]{2017PASA...34...58E,2018MNRAS.479...75S} predictions for unbound companions to the first supernova in a binary \citep{2023MNRAS.522.2029C}. The absolute magnitudes are corrected for the line-of-sight extinction, at the inferred distance of each object, using the Bayestar 3D dust map \citep{2019ApJ...887...93G}, the effective wavelength of the {\em Gaia} $G$-filter, assuming R$_{V}=3.1$, and a Fitzpatrick extinction law \citep{1999PASP..111...63F}. We again emphasise that on this sight-line, the maximum distance and extinction ($\sim$4\,kpc with A$_{V} \sim 3$) mean that effectively all OB stars should be brighter than {\em Gaia}'s G$\sim$21 limiting magnitude. 

Although the two objects in the right-hand panel of Figure \ref{figure:comp} have distances and past trajectories consistent with \sgr, the lack of an obvious SNR or stellar cluster associated with \sgr\ puts their nature as potential runaways associated with \sgr\ into doubt. To further verify this, we rotate the proper motion vector of \sgr\ by 90, 180 and 270\,deg, repeating the analysis. Objects whose past trajectories intersect \sgr\ under these conditions are shown in the left-hand panel of Figure \ref{figure:comp} by grey lines. The `real' matches, shown in colour, do not stand out from the population of random past associations. This is consistent with the low association probabilities we calculated for the two objects of $\lesssim 10^{-3}$. Hence we cannot confidently identify any likely runaway candidate associated with \sgr\, nor any clear SNR or cluster of origin. This is a stronger statement that can be made for objects closer to the Galactic centre, where even bright stars and star formation may be missed due to high extinction, and numerous smaller star forming regions are likely present.

\subsection{Association with other supernova remnants and star forming regions}
We now consider the possibility that we have missed other SNRs or star forming regions that are plausibly associated with \sgr. 

In addition to HB9, another SNR, G160.1-1.1, is worth consideration. Only a small arc is visible \citep[see Figure 10 of][]{2014A&A...566A..76G}, but if extrapolated into a circle, the radius would be large enough to reach \sgr. The faint nebulosity at (160${\degr}$,1${\degr}$) in the right-hand panel of Figure \ref{figure:location} might plausibly be associated with G160.1-1.1. However, the proper motion of \sgr\ does not trace back to the centre of this SNR either, and \sgr\ would have to be far older (and the SNR far older, given the slow proper motion of \sgr) than estimated for this scenario to work. We therefore deem it unlikely that G160.1-1.1 is associated with \sgr.

The wider field around \sgr\ is notable for two star forming complexes, Sh217 and Sh219. They appear
to lie at much larger distance that HB9, probably in the outer arm at $\sim 5$kpc, suggesting that there is some massive star formation at these distances. However, both of these regions have larger projected offsets from the location of \sgr\ than HB9, and are offset in directions well away from the proper motion vector. We therefore rule these out as the origin of the SGR.

\sgr\ lies close to the Galactic plane, in the region surveyed by both IPHAS \citep{2005MNRAS.362..753D}, 
and the UKIRT Infrared Deep Sky Survey (UKIDSS) Galactic Plane Survey \citep[GPS][]{2008MNRAS.391..136L}. This offers the opportunity to survey the environs of \sgr\ at higher resolution, and to greater depth than is possible
via the Digital Sky Survey (DSS) or 2MASS. In addition, the H$\alpha$ survey is ideal for locating star forming regions, or
supernova remnants which could represent the birthplace of the SGR. We show the past trajectory of \sgr\, overlaid on IPHAS H$\alpha$ imaging, in Figure \ref{figure:location}. Within a giant molecular cloud the density may be of order $10^2 - 10^3$ cm$^{-3}$,
compared to typical ISM densities of order 1 cm$^{-3}$. 
In such cases, the SNR rapidly sweeps up its
own mass and hence is confined within a small volume, as it cannot expand significantly during the
free streaming phase. For example, the compact supernovae remnants observed in M82
are thought to be only 0.6-4 pc in radius \citep{2001ApJ...558L..27C}. This corresponds to 
$\sim$60--400\,$d_{2{\rm kpc}}^{-1}$ \arcsec. Inspection of the available imaging shows no sign of such a cluster or compact SNR within
the region constrained by the observed limits on the proper motion combined with the spin-down age.

It is possible, of course, that the progenitor star of \sgr\ was not in a cluster at the time of its
explosion. Under the most popular variants of models for SGR production this seems unlikely, since 
the progenitor is expected to be a massive star \citep[e.g. the suggested
progenitor masses for magnetars are in the range $\sim$15-50\,M${\odot}$,][]{2006ApJ...636L..41M,2008MNRAS.386L..23B,2009ApJ...707..844D,2017ApJ...846...13B}, which is unlikely to travel large transverse distances in its short lifetime \citep[although see e.g.][]{2011MNRAS.414.3501E}. In any case, even if the progenitor exploded outside a cluster, we would expect to easily locate the associated supernova remnant. The ISM density within the Galactic disk is sufficiently high to produce a bright shock, though low enough to enable the SNR to grow to a moderate size in $\sim 20$\,kyr, as is the case for HB9.

\section{Interpretation}
We conclude that with the available data there is no sign of either a SNR or young cluster consistent with the birth site of \sgr. SGR\,0501+4516 is at an unusual Galactic location with respect to most other magnetars, lying in the Galactic anticentre direction, towards the Perseus arm at $\sim$2\,kpc and the Outer arm at $\sim$5\,kpc. Only SGR\,0418$+$5729, 1E\,2259$+$586 and 4U\,0142$+$61 have a comparable location within the Galaxy \citep{2010ApJ...711L...1V}. However, 1E\,2259$+$586 has a robust SNR association \citep{2013ApJ...772...31T}, SGR\,0418$+$5729 has a much older characteristic age of 36\,Myr \citep{2014ApJS..212....6O}, and 4U\,0142$+$61 likely lies more distant and behind higher extinction. \sgr\ therefore remains one of the youngest, likely nearest and least attenuated magnetars in the Galactic population, yet it lacks a clear birth site or associated SNR. This makes \sgr\ the best known candidate for a magnetar formed in a process other than massive star core-collapse, or might suggest that it had an otherwise unusual progenitor and/or an age which is much larger than expected. We now examine each of these possibilities in turn.

\subsection{Uncertainties on the age of SGR\,0501+4516}\label{sec:oldmagnetar}
One possibility is that \sgr\ is much older than its characteristic age implies. The characteristic age
relies on measurements of the period derivative at the current time. To provide an 
accurate measurement of the age, it requires that spin down has proceeded in a relatively 
uniform manner, and that the initial period was much smaller than the currently observed one. It is possible that, associated with outburst activity, a magnetar exhibits a period of accelerated spin-down, resulting in a larger value of $\dot{P}$ than magnetic braking alone would generate and causing the characteristic age
of the neutron star to be underestimated. If this is the case, the true age of the magnetar may be
order of magnitude larger than the characteristic age. However, even within the region traversable by \sgr\ in the past several tens of Myr, there are only two SNRs (HB9 and G160.1-1.1), both of which appear inconsistent with the proper motion vector as discussed above, although at ages of $10^5$ years SNRs can become extremely diffuse and essentially invisible to observations, particularly if overlapping in projection with bright structures (such as HB9). Interestingly, \citet{2007ApJ...662.1198H}, exploiting the well-determined proper motion of AXP\,XTEJ1810-197, found no compelling candidate birth sites for an age of less than $\sim10^5$ yrs (although several SGRs, e.g. SGR\,1806-20 and SGR\,1900+14, were born in nearby bright clusters, as noted above). In these
cases, the high ISM pressure inside the clusters could confine, and effectively mask, the presence of a SNR, explaining the lack of an apparent association. Indeed, if the lifetimes of magnetars can be significantly longer than the canonical $10^4$ years this would help to ameliorate concerns over the rate of magnetar production (given that $\sim30$ confirmed or candidate magnetars are now known). 

By contrast, the AXP\,1E\,2259+586 lies within the supernova remnant CTB\,109, but has a characteristic age approximately 20 times the SNR age, suggesting that sometimes the characteristic age can significantly overestimate the age of the magnetar (e.g. because the magnetic field has decayed with time, and magnetic braking was stronger at earlier times). Although deviations between $P/2\dot{P}$ and true ages may plausibly operate in both directions, recent works favour the interpretation that magnetar (and pulsar) characteristic ages are typically overestimates of their true ages \citep{2019MNRAS.487.1426B,2021ApJ...913L..12M,2022MNRAS.517.3008P,2024ApJ...976..228R}.

\subsection{SGR0501+4516 as the product of a low-ejecta mass supernova}
If the true age of \sgr\ is instead comparable to its characteristic age, and it did form in a core-collapse event, then a low supernova ejecta mass may explain the lack of a bright supernova remnant (as well as the low peculiar velocity). As discussed in Section \ref{sec:birthsite}, the ISM density on the sight-line to \sgr\ is broadly conducive to producing moderately large, moderately bright supernova remnants such as HB9. High ISM densities would confine the SNR (making it bright but small in extent), low ISM densities allow a remnant to rapidly expand, but the surface brightness will be low. If \sgr\ did form in a supernova at the approximate location implied by the left panel of Figure \ref{figure:location}, the apparent lack of a SNR there may be due to a low ejecta mass \citep[e.g.][]{2015MNRAS.451.2123T}, producing a remnant which is smaller, fainter and more rapidly fading than a typical core-collapse SNR. Low ejecta masses can be produced by stripped (or ultra-stripped) stars, which are likely produced through stripping by a binary companion, and may explode as electron capture supernovae \citep{1987ApJ...322..206N}. Such a scenario may produce small natal kicks \citep[e.g.][]{2016MNRAS.461.3747B}. Progenitor masses for magnetars based on the ages of associated stellar clusters and supernova remnant modelling yield a wide range of $\sim$15--50\,M$_{\odot}$ \citep{2006ApJ...636L..41M,2008MNRAS.386L..23B,2009ApJ...707..844D,2017ApJ...846...13B}. The seemingly low natal kick velocity and lack of other massive stars and clusters in the vicinity of \sgr\ makes a similar interpretation for this object challenging.

\subsection{SGR0501+4516 as the product of a binary neutron star merger}
Magnetars can also emerge from the merger of two low-mass neutron
stars (NSs) in a binary system when the remnant mass post-merger falls
below the Tolman–Oppenheimer–Volkoff (TOV) limit, $M_{\rm TOV}$, which
defines the maximum mass a non-rotating NS can have while remaining
stable against gravitational collapse \citep{Oppenheimer1939}. If the
mass of the newly formed NS is below this threshold, the remnant can
reach hydrostatic equilibrium without requiring additional support
from rapid rotation or magnetic fields to counteract gravitational
forces.

General Relativistic Magnetohydrodynamic (GRMHD) simulations by
various independent groups have demonstrated that, when two NSs merge,
the resulting magnetic field can be amplified by several orders of
magnitude. This amplification is driven by a combination of magnetic
winding due to differential rotation and instabilities such as the
Kelvin-Helmholtz instability and magneto-rotational instabilities
\citep{Giacomazzo2013, Kiuchi2015, Giacomazzo2015, Ciolfi2017, Palenzuela2022, Aguilera2023, Kiuchi2024}. If the merger remnant has a low enough mass to remain stable, it is
therefore very likely to acquire a magnetic field in the magnetar range. The remnant may also receive a recoil kick through anisotropic mass loss and gravitational wave emission \citep{2023PhRvD.108j3023K}.

Magnetars formed through the binary NS merger channel are expected to
reside outside of star-forming regions due to the long delay times
often associated with binary NS mergers and the kicks imparted by
supernovae to each progenitor NS in the binary. These characteristics
align with observed properties of short gamma-ray bursts,
which are frequently found in low star-formation regions and are
known to be associated with binary NS mergers (see e.g. \citealt{Berger2014}
for a review).

If \sgr\ is indeed a stable, highly magnetized NS resulting from
the merger of two NSs, an interesting question is the expected
contribution of such merger products to the overall magnetar
population. The fraction of binary NS mergers that produce a stable NS
remnant is highly sensitive to both the NS mass function and the
equation of state (EoS) of neutron star matter. Studies by
\citet{Piro2017}, assuming a Gaussian distribution for the NS mass
function based on the observed Galactic NS population, surveyed the
remnant populations for several representative EoS models. They found
that the fraction of mergers producing stable NSs could range from
negligible in an EoS like H4 \citep{Lackey2006} to a few
percent in a stiffer EoS like APR4 \citep{Akmal1998}, and
even the majority of cases for very stiff EoS models like SHT
\citep{Shen1998,Shen2011}. Thus, the discovery of a stable magnetar
from a binary NS merger may provide constraints on the EoS if this
magnetar population can be accurately identified and quantified.

\subsection{\sgr\ as the product of accretion or merger-induced collapse}
A final alternative is that \sgr\ formed via the accretion or merger-induced collapse
of a magnetic white dwarf \citep[e.g.][]{1979wdvd.coll...56N,1991ApJ...367L..19N,1999ApJ...516..892F,2006MNRAS.368L...1L,2007MNRAS.380..933Y,2008MNRAS.385.1455M,2013A&A...558A..39T,2019MNRAS.484..698R,2022MNRAS.509.6061A,2025ApJ...978L..38C}. This may occur via mass transfer within a
binary, or the merger of two white dwarfs (WDs), at least one of which is magnetic and with an ONeMg composition. In these scenarios, when the mass of the WD exceeds the Chandrasekhar mass, it undergoes collapse to a neutron star.
In the most simple model, it is assumed that the magnetic flux is conserved during collapse ($BR^2 =$ constant, where $B$ is the magnetic field strength and $R$ the radius), and hence, for a change in radius of a factor 500--1000 (typical for a WD and neutron star) the increase
in the magnetic field strength can be a factor of $10^6$. Highly magnetic WDs, with
$B$-fields of $10^8-10^9$\,G, could then become magnetars \citep{2006MNRAS.368L...1L}.

This channel is appealing for \sgr\ since it is likely that minimal 
mass is ejected in such a merger \citep[e.g.][]{1999ApJ...516..892F}. The mean mass of a WD is $\sim 0.5-0.6$ M$_{\odot}$ \citep[e.g.][]{2024A&A...690A..68S}, and 
so the majority of WD-WD systems which exceed the Chandrasekhar mass may well do so only marginally.
In this case the majority of the mass is likely to remain in the newly formed neutron star, leading to low mass ejection and energy, and hence a faint remnant. Thus, in this model the lack of either a supernova remnant, or a young cluster, can naturally be explained. Furthermore, the low peculiar transverse velocity of \sgr\ is comparable with the peculiar velocities of white dwarfs \citep{2022A&A...658A..22R,2022MNRAS.512.6201M}. This is the qualitative expectation in an accretion or merger-induced collapse scenario, where low ejecta masses \citep[fractions of a Solar mass, e.g.][]{2025ApJ...978L..38C} would likely produce weak natal kicks.

It is reasonable to wonder if the rates of such channels are plausible. Estimates for the core-collapse magnetar formation rate, based on the observed Galactic population size and their characteristic ages, can be as high as $>$50\% of all core-collapse neutron stars being born as magnetars \citep{2015MNRAS.454..615G,2019MNRAS.487.1426B}. However, such estimates are heavily influenced by observational biases and are highly uncertain. On the WD-WD side, up to $\sim$20\% of high-mass WDs are thought to be merger products, and the total Galactic rate of WD-WD mergers (of all masses) is estimated at $\sim$0.1\,yr$^{-1}$ \citep{2020ApJ...891..160C}. However, to produce a magnetar through this channel, we also require (i) that the final mass exceeds the Chandrasekhar mass, (ii) that the chemical composition is conducive to neutronisation rather than a runaway thermonuclear reaction, thus avoiding a type Ia SN, (iii) that the progenitor WD was magnetic. Mergers of WD binaries above the Chandrasekhar mass are expected at the level of $10^{-3}$\,yr$^{-1}$ \citep{2001A&A...365..491N,2006MNRAS.368L...1L}. Since $\gtrsim$10\% of WDs are magnetic \citep{2017MNRAS.467.4970H}, and assuming that these more massive WDs have the ONeMg composition required for neutronisation rather than thermonuclear runaway, the Galactic rate of magnetars from merger-induced collapse is therefore around $10^{-4}$\,yr$^{-1}$. Estimates for magnetar formation from the single-degenerate AIC channel are similar, or slightly lower \citep[$3\times10^{-5}$\,yr$^{-1}$,][]{2022MNRAS.509.6061A}. Taking the lower end of the core-collapse magnetar formation rate \citep[$\sim10^{-3}$\,yr$^{-1}$][]{2019MNRAS.487.1426B}, and the estimated rate of magnetars from merger and accretion-induced collapse ($\sim10^{-4}$\,yr$^{-1}$), we could plausibly expect as many as 10\% of Galactic magnetars to have a non-core-collapse origin. 

Although these rates are uncertain, they suggest that under the assumption that some WD-WD merger leads to collapse to a NS, rather than SN Ia, it is reasonable to expect a non-negligible fraction of SGR and AXPs to originate via this route, and that locating objects such as \sgr\ would not be unexpected in these circumstances. Currently only two magnetars out of a Galactic population of $\sim$30 objects - \sgr\ and potentially 4U\,0142 - are known to have young characteristic ages and relatively clear sight-lines, such that it is surprising that no birth site has been identified. Out of these, \sgr\ is the nearest and least attenuated, and thus offers the best constraint.

\subsection{Other evidence for delayed magnetar formation channels}
An intriguing observation is that a subset of Fast Radio
Bursts (FRBs) has been detected in galaxies with no apparent star
formation activity, similar to the environment of \sgr. FRBs are widely believed to be associated with magnetars, particularly following the observation of FRB-like bursts from Galactic magnetar SGR\,1935$+$2154
\citep[][]{Margalit2018,2019ApJ...886..110M,2020Natur.587...59B,2020Natur.587...54C,2020ApJ...898L..29M,2021NatAs...5..414K}. Observationally, around 5\% of FRBs have been linked with ancient stellar populations \citep[by residing in massive quiescent galaxies, and in one case a globular cluster,][]{2021ApJ...917L..11K,2022Natur.602..585K,2023ApJ...954...80G,2023ApJ...950..175S,Eftekhari2024,2025ApJ...979L..21S}. A longer radio flash has also been linked with a (possibly magnetar-driven) compact-merger gamma-ray burst \citep{2024MNRAS.534.2592R,2024ApJ...973L..20S}. This suggests that at least a fraction of these sources may have formed
through delayed channels such as binary NS mergers or white dwarf accretion induced collapse, rather than recent star formation events. As the sample sizes for both FRBs and SGRs continue to grow, comparative analysis of magnetars across formation channels will provide insights into their population fractions, potentially allowing further constraints on the mechanisms producing these highly magnetized objects.

\section{Conclusions}
We have presented NIR observations of the counterpart of \sgr, demonstrating that the NIR flux of the counterpart broadly follows the X-ray variability. Additionally, the long temporal baseline of our {\em HST} observations enabled us to measure the proper motion of the SGR, which corresponds to a low transverse peculiar velocity. The direction of the proper motion vector rules out an association with the nearby supernova remnant HB9, and there is no star formation or reasonable alternative SNR located close to the SGR. This suggests that either (i) \sgr\ is much older than typically anticipated for a magnetar, in contrast with recent modelling which indicates that SGR spin-down ages are overestimates, (ii) that the progenitor was a massive star but had a low supernova ejecta mass, or (iii) that it may have formed via a route that does not (directly) involve massive star core-collapse. While not conclusive, \sgr\ therefore represents the best candidate currently known for a Galactic magnetar formed through the merger of two low mass neutron stars, or the accretion-induced collapse of a magnetic white dwarf.

\begin{acknowledgements} 
The authors thank Jos de Bruijne, David Green, and David O'Neill for helpful discussions, and the anonymous referee for their careful consideration of the manuscript.

AAC acknowledges support through the European Space Agency (ESA) research fellowship programme.
JDL acknowledges support from a UK Research and Innovation Fellowship(MR/T020784/1).
VSD is supported by STFC.
NRT is supported by STFC Consolidated grant ST/W000857/1.
NR is supported by the European Research Council (ERC) via the Consolidator Grant “MAGNESIA” (No. 817661) and the Proof of Concept "DeepSpacePulse" (No. 101189496), by the Catalan grant SGR2021-01269, by the Spanish grant  PID2023-153099NA-I00, and by the program Unidad de Excelencia Maria de Maeztu CEX2020-001058-M.
DS acknowledges support from the Science and Technology Facilities Council (STFC, grant numbers ST/T007184/1, ST/T003103/1, ST/T000406/1 and ST/Z000165/1).
The material is based upon work supported by NASA under award number 80GSFC24M0006.

Observations analysed in this work were taken by the NASA/ESA Hubble
Space Telescope under programs 12306, 12672 and 16019 (PI: Levan). This work made use of data from the European Space Agency
(ESA) mission {\em Gaia} (\url{https://www.cosmos.esa.int/{\em Gaia}}), processed
by the {\em Gaia} Data Processing and Analysis Consortium (DPAC, \url{https://www.cosmos.esa.int/web/{\em Gaia}/dpac/consortium}). Funding for the DPAC has been provided by national institutions, in particular the institutions participating in the {\em Gaia} Multilateral Agreement.

This paper makes use of data obtained as part of the IGAPS merger of the IPHAS and UVEX surveys (\url{www.igapsimages.org}) carried out at the Isaac Newton Telescope (INT). The INT is operated on the island of La Palma by the Isaac Newton Group in the Spanish Observatorio del Roque de los Muchachos of the Instituto de Astrofisica de Canarias. All IGAPS data were processed by the Cambridge Astronomical Survey Unit, at the Institute of Astronomy in Cambridge. The uniformly-calibrated bandmerged IGAPS catalogue was assembled using the high performance computing cluster via the Centre for Astrophysics Research, University of Hertfordshire.

The United Kingdom Infrared Telescope is operated by the Joint
Astronomy Centre on behalf of the Science and Technology Facilities
Council of the U.K.  We thank Nancy Levison for awarding DDT observations
with Gemini, and the ING group for their assistance with our observations
at the WHT. The WHT is operated on the island of La Palma by
the Isaac Newton Group in the Spanish Observatorio del Roque de los
Muchachos of the Instituto de Astrofísica de Canarias. Based on
observations obtained at the Gemini Observatory, which is operated by
the Association of Universities for Research in Astronomy, Inc., under
a cooperative agreement with the NSF on behalf of the Gemini
partnership: the National Science Foundation (United States), the
Science and Technology Facilities Council (United Kingdom), the
National Research Council (Canada), CONICYT (Chile), the Australian
Research Council (Australia), Ministério da Ciência e Tecnologia
(Brazil) and SECYT (Argentina).  This publication makes use of data
products from the Two Micron All Sky Survey, which is a joint project
of the University of Massachusetts and the Infrared Processing and
Analysis Center/California Institute of Technology, funded by the
National Aeronautics and Space Administration and the National Science
Foundation.

This work made use of v2.2.1 of the Binary Population and Spectral Synthesis (BPASS) models as described in \citet{2017PASA...34...58E} and \citet{2018MNRAS.479...75S}. This work has made use of {\sc ipython} \citep{2007CSE.....9c..21P}, {\sc numpy} \citep{2020Natur.585..357H}, {\sc scipy} \citep{2020NatMe..17..261V}; {\sc matplotlib} \citep{2007CSE.....9...90H}, Seaborn packages \citep{Waskom2021} and {\sc astropy} (\url{https://www.astropy.org}) a community-developed core Python package for Astronomy \citep{astropy:2013, astropy:2018}. We have also made use of the python modules {\sc dustmaps} \citep{2018JOSS....3..695M} and {\sc extinction} \citep{barbary_kyle_2016_804967}. 

\end{acknowledgements}

%\begin{thebibliography}{}
\bibliographystyle{aa} % style aa.bst
\bibliography{IR_0501} % your references Yourfile.bib
%\end{thebibliography}{}

%\begin{appendix} %First appendix

%\end{appendix}

%%%% End of aa.dem
\end{document}